\documentclass[11pt,a4paper]{article} 
\pdfoutput=1
\usepackage{jheppub,hyperref,mdwlist}
\usepackage{amsmath}
\usepackage{graphicx}
\usepackage{subfigure}
\usepackage{amssymb}
\usepackage{xcolor}
\usepackage{multirow}
\usepackage{cancel}
\usepackage{color}
\usepackage{listings}
\usepackage{xcolor}
\usepackage{float}
\usepackage{slashed}

\usepackage{amsmath}
\usepackage{wrapfig}

\newcommand{\be}{\begin{equation}}
\newcommand{\ee}{\end{equation}}
\newcommand{\beq}{\begin{equation}}
\newcommand{\eeq}{\end{equation}}
\newcommand{\bea}{\begin{eqnarray}}
\newcommand{\eea}{\end{eqnarray}}
\newcommand{\besp}{\begin{equation}\begin{split}}
\newcommand{\eesp}{\end{split}\end{equation}}

\newcommand{\nn}{\nonumber}

\newcommand{\Br}{\text{Br}}

\newcommand{\tabincell}[2]{\begin{tabular}{@{}#1@{}}#2\end{tabular}}
\newcommand{\Eq}[1]{Eq.~(\ref{#1})}

\newcommand{\Dfbd}{\mathord{\buildrel{\lower3pt\hbox{$\scriptscriptstyle\leftrightarrow$}}\over {D}_{\mu}}}
\newcommand{\ave}[1]{\left\langle #1\right\rangle}

\def\hc{{\rm h.c.}}

\def\mL{\mathcal{L}}

\def\mO{\mathcal{O}}

\def\mT{\mathcal{T}}

\def\Z{\mathbb{Z}}

\def\0{\textbf{0}}
\def\1{\textbf{1}}
\def\2{\textbf{2}}
\def\3{\textbf{3}}
\def\4{\textbf{4}}
\def\5{\textbf{5}}
\def\6{\textbf{6}}
\def\7{\textbf{7}}
\def\8{\textbf{8}}
\def\9{\textbf{9}}

\begin{document}

\title{Lepton-mediated electroweak baryogenesis, gravitational waves and the $4\tau$ final state at the collider}

\author[a]{Ke-Pan Xie,}
\affiliation[a]{Center for Theoretical Physics, Department of Physics and Astronomy, Seoul National University, Seoul 08826, Korea}

\emailAdd{kpxie@snu.ac.kr}

\abstract{An electroweak baryogenesis (EWBG) mechanism mediated by $\tau$ lepton transport is proposed. We extend the Standard Model with a real singlet scalar $S$ to trigger the strong first-order electroweak phase transition (SFOEWPT), and with a set of leptophilic dimension-5 operators to provide sufficient CP violating source. We demonstrate this model is able to generate the observed baryon asymmetry of the universe. This scenario is experimentally testable via either the SFOEWPT gravitational wave signals at the next-generation space-based detectors, or the $pp\to h^*\to SS\to 4\tau$ process (where $h^*$ is an off-shell Higgs) at the hadron colliders. A detailed collider simulation shows that a considerable fraction of parameter space can be probed at the HL-LHC, while almost the whole parameter space allowed by EWBG can be reached by the 27 TeV HE-LHC.
}

\maketitle
\flushbottom

\section{Introduction}

Cosmological observations show that there is an imbalance between the abundance of matter and antimatter in our universe. This is also known as the baryon asymmetry of the universe (BAU), which can be described by the baryon-to-entropy ratio~\cite{Tanabashi:2018oca}
\be
\eta_B\equiv \frac{n_B}{s}=[0.82\sim0.94]\times10^{-10}.
\ee
Creating the BAU requires three Sakharov conditions~\cite{Sakharov:1967dj,Kuzmin:1970nx,Ignatiev:1978uf}: i) baryon number violation, ii) C and CP violation, and iii) departure from equilibrium. The Standard Model (SM) satisfies the first condition [via the electroweak (EW) sphaleron] but fails in the last two: the CP violating (CPV) phase [from the CKM matrix] is too tiny, and the SM EW phase transition (EWPT) is a smooth crossover~\cite{Rummukainen:1998as}. Therefore, the observed BAU is a clear evidence for physics beyond the SM (BSM).

There have been a lot of BSM mechanisms explaining BAU~\cite{Bodeker:2020ghk}, and this paper will focus on EW baryogenesis (EWBG)~\cite{Morrissey:2012db,Cline:2006ts,Trodden:1998ym}. In EWBG, the out-of-equilibrium environment is provided by a strong first-order EWPT (SFOEWPT), during which bubbles containing EW broken phase nucleate and expand in the EW symmetric background. The CPV source on the bubble wall seeds a net density for the left-handed fermions in front of the wall when the particles collide with the bubbles. This ``chiral asymmetry'' then diffuses into the EW symmetric phase where it is converted to a baryon asymmetry by the EW sphaleron process. The net baryon number is swept into the expanding bubbles and then frozen, because the EW sphaleron is strongly suppressed inside the bubble. The bubbles eventually fill up the whole universe, completing the SFOEWPT and leaving the observed BAU. EWBG is an attractive scenario as it typically involves BSM physics around TeV scale, thus can be tested at current or future colliders~\cite{Arkani-Hamed:2015vfh}. In addition, the phase transition gravitational waves (GWs) might be detectable at the future space-based detectors~\cite{Mazumdar:2018dfl}.

Most literatures focus on EWBG mediated by quark transport, in particular by top transport since the large top Yukawa ($y_t\approx1$) enhances the CPV source and hence the BAU~\cite{Joyce:1994fu,Joyce:1994zt,Fromme:2006wx} (for recent studies see~\cite{Jiang:2015cwa,Bruggisser:2018mus,Chao:2019smr,Ellis:2019flb,DeCurtis:2019rxl,Cline:2020jre,Xie:2020bkl})\footnote{There are also EWBG mechanisms mediated by bottom~\cite{Modak:2018csw,Modak:2020uyq} or charm~\cite{Bruggisser:2018mrt} quark transport.}. However, the top-mediated EWBG is not as efficient as expected, due to three reasons summarized in Ref.~\cite{deVries:2018tgs} as follows. First, because of the strong QCD interaction with the plasma, the diffusion of quark chiral asymmetry to the EW symmetric phase is inefficient~\cite{Joyce:1994bi}. Second, the CPV in top sector has been stringently constrained by the electron electric dipole moment (EDM) experiments~\cite{Cirigliano:2016nyn}\footnote{Actually, once the EDM measurements are taken into account, the top-mediated EWBG within the SM effective field theory (EFT) framework can only give a BAU that is 2 orders of magnitude smaller than the observed value~\cite{deVries:2017ncy}. However this constraint can be relaxed if we consider new light BSM degrees of freedom. For example, the SM plus a real singlet scalar is able to generate the observed BAU and compatible with the EDM experiments~\cite{Espinosa:2011eu,DeCurtis:2019rxl}, especially when there is a $\Z_2$ symmetry for the singlet~\cite{Xie:2020bkl}.}. Third, the top Yukawa as well as the strong sphaleron process tend to washout the chiral asymmetry, significantly decreasing the generated BAU~\cite{Giudice:1993bb,Tulin:2011wi}.

Those three disadvantages may be overcome in a lepton-mediated EWBG~\cite{deVries:2018tgs}: first, leptons are easier to diffuse in the plasma~\cite{Joyce:1994bi}; second, the EDM constraints for muon and tau are much weaker~\cite{Brod:2013cka}; third, the leptons are free from the washout effect of strong sphaleron. Therefore, even though the CPV source from the lepton sector is typically smaller than that from the top sector (for example the tau Yukawa $y_\tau\approx0.01\ll y_t$), a lepton-mediated EWBG is still possible to account for the observed BAU. Indeed, Refs.~\cite{Joyce:1994zn,Chung:2008aya,Chung:2009cb,Chiang:2016vgf,Guo:2016ixx} have shown that a successful $\tau$-mediated EWBG can be realized, if $y_\tau$ is enhanced in BSM models. Recently, Refs.~\cite{deVries:2018tgs,Fuchs:2020uoc} study the CPV source from SM EFT dimension-6 operators, and point out that even a SM-like $y_\tau$ is able to realize EWBG.\footnote{See Ref.~\cite{Fuchs:2019ore} for a $\mu$-mediated EWBG, and Ref.~\cite{Fernandez-Martinez:2020szk} for a neutrino-mediated EWBG.}

In this work, we propose a novel $\tau$-mediated EWBG model based on the real singlet scalar extension of SM (xSM). The CPV source comes from the leptophilic dimension-5 operator $(1/\Lambda)S\bar\ell_L H\tau_R$, with $\Lambda$ being the cutoff scale, $S$ the singlet, $H=\left(G^+,(h+iG^0)/\sqrt{2}\right)^T$ the Higgs doublet, $\ell_L=(\nu_{\tau L},\tau_L)^T$ the third generation lepton doublet and $\tau_R$ the right-handed tau. We will show that such a model can explain today's BAU with a SM-like $y_\tau$ in Section~\ref{sec:EWBG}, where we also discuss the possibility of detecting the phase transition GWs at the future space-based detectors. Section~\ref{sec:4tau} is devoted to collider phenomenology, in which we demonstrate our scenario can be efficiently probed by the $pp\to h^*\to SS\to \tau^+\tau^-\tau^+\tau^-$ process at the LHC and future 27 TeV HE-LHC. Finally, we conclude in Section~\ref{sec:conclusion}.

\section{Lepton-mediated EWBG}\label{sec:EWBG}

\subsection{SFOEWPT and GWs}\label{sec:GWs}

The SFOEWPT of xSM has been extensively studied~\cite{McDonald:1993ey,Profumo:2007wc,Espinosa:2011ax,Cline:2012hg,Alanne:2014bra,Vaskonen:2016yiu,Huang:2018aja,Cheng:2018ajh,Alanne:2019bsm,Gould:2019qek,Carena:2019une}, and we needn't to repeat all those results here. For simplicity, we assume the following tree level potential
\be\label{general_V0}
V_0=\mu_H^2|H|^2+\frac{\mu_S^2}{2}S^2+\lambda_H|H|^4+\frac{\lambda_S}{4}S^4+\lambda_{HS}|H|^2S^2,
\ee
which is symmetric under the $\Z_2$ transformation $S\to-S$. Under the unitary gauge,
\be\label{sim_V0}
V_0=\frac{\mu_H^2}{2}h^2+\frac{\mu_S^2}{2}S^2+\frac{\lambda_H}{4}h^4+\frac{\lambda_S}{4}S^4+\frac{\lambda_{HS}}{2}h^2S^2.
\ee
The bounded below condition for this potential is $\lambda_H>0$, $\lambda_S>0$, and $\sqrt{\lambda_H\lambda_S}+\lambda_{HS}>0$. We are interested in the parameter space with~\cite{Bian:2019kmg}
\be
\mu_H^2<0,\quad \mu_S^2<0,\quad \lambda_H\mu_S^2>\lambda_{HS}\mu_H^2,\quad \lambda_S\mu_H^2>\lambda_{HS}\mu_S^2,\quad -\frac{\mu_H^4}{\lambda_H}<-\frac{\mu_S^4}{\lambda_S},
\ee
in which $V_0$ has two local minima $(v,0)$ and $(0,w)$ with
\be\label{tree-level_1}
v=\sqrt{-\mu_H^2/\lambda_H},\quad w=\sqrt{-\mu_S^2/\lambda_S},
\ee
and the former is the global minimum, i.e. the vacuum. Expanding potential around the vacuum gives physical masses of the scalars
\be\label{tree-level_2}
M_h^2=-2\mu_H^2,\quad M_S^2=\mu_S^2+\lambda_{HS}v^2.
\ee
In other words, the measured values $M_h=125$ GeV and $v=246$ GeV have fixed $\mu_H^2$ and $\lambda_H$, and we only have three free parameters in \Eq{sim_V0}.

At one-loop level, $V_0$ becomes
\be\label{V1}
V_1=V_0+V_{\rm CW}+V_{\rm CT},
\ee
where $V_{\rm CW}$ is the Coleman-Weinberg potential, and $V_{\rm CT}$ is the counter term. At finite temperature, the scalar potential receives thermal corrections,
\be\label{VT}
V_T=V_1+\tilde V_T+V_{\rm daisy},
\ee
where $\tilde V_T$ is the one-loop thermal integral correction, while $V_{\rm daisy}$ is the daisy resummation term. The detailed expressions for $V_{\rm CW}$, $V_{\rm CT}$, $\tilde V_T$ and $V_{\rm daisy}$ are given in Appendix~\ref{app:one-loop}. Taking only the leading $T^2$ terms~\cite{Dolan:1973qd}, $V_T$ can be approximated as
\be\label{VT_sim}
V_T\approx\frac{1}{2}\left(\mu_H^2+c_HT^2\right)h^2+\frac12\left(\mu_S^2+c_S T^2\right)S^2+\frac{\lambda_H}{4}S^4+\frac{\lambda_S}{4}S^4+\frac{\lambda_{HS}}{2}h^2S^2,
\ee
where 
\be
c_H=\frac{3g^2+g'^2}{16}+\frac{y_t^2}{4}+\frac{\lambda_H}{2}+\frac{\lambda_{HS}}{12},\quad c_S=\frac{\lambda_S}{4}+\frac{\lambda_{HS}}{3},
\ee
with $g^{(\prime)}$ and $y_t$ being the EW gauge couplings and top Yukawa, respectively.

The thermal potential in \Eq{VT_sim} can realize a two-step phase transition in which a second-order phase transition first occurs along the $S$-axis, and then a first-order EWPT happens through the vacuum decay between the $S$- and $h$- axes, i.e.
\be
\left(\ave{h}=0,\ave{S}=0\right)\xrightarrow[\rm step]{\rm First}\left(\ave{h}=0,\ave{S}\neq0\right)\xrightarrow[\rm step]{\rm Second}\left(\ave{h}\neq0,\ave{S}=0\right),
\ee
as the temperature decreases. For the first step, if there were an exact $\Z_2$ symmetry for $S$ in $V_T$, then the probability of transition to $+S$ and $-S$ directions would be equal, leaving the domain wall problem. To avoid this issue, we assume a small $\Z_2$-breaking is present, such that the transition along the $+S$ direction is energetically preferred~\cite{Espinosa:2011eu}.

The second step is a first-order EWPT, whose necessary condition is two degenerate vacua separated by a barrier at the critical temperature $T_c$. For the polynomial potential in \Eq{VT_sim}, this condition is equivalent to~\cite{Bian:2019kmg} 
\be\label{degenerate_V}
\frac{c_S}{c_H}<\frac{\mu_S^2}{\mu_H^2}<\frac{\sqrt{\lambda_S}}{\sqrt{\lambda_H}}<\frac{\lambda_{HS}}{\lambda_H}.
\ee
The sufficient condition for a first-order EWPT is the onset of nucleation, which is resolved from the equality of vacuum decay rate and the universe expansion rate at the nucleation temperature $T_n$,
\be\label{Linde}
T_n^4e^{-S_3(T_n)/T_n}\approx H^4(T_n),
\ee
with $S_3(T)$ being the Euclidean action of the $O(3)$-symmetric bounce solution~\cite{Linde:1981zj}, and $H(T)$ the Hubble constant. For a radiation-dominated universe, \Eq{Linde} is approximately
\be\label{n_condition}
\frac{S_3(T_n)}{T_n}\sim140.
\ee
for a $T_n$ around the EW scale~\cite{Quiros:1999jp}. For the EW sphaleron process to be suppressed inside the bubble, we further require~\cite{Moore:1998swa,Zhou:2019uzq}
\be\label{S_condition}v_n/T_n\gtrsim1,
\ee
which is the definition for a strong transition, namely a SFOEWPT.

\begin{figure}
\centering
\subfigure{
\includegraphics[scale=0.55]{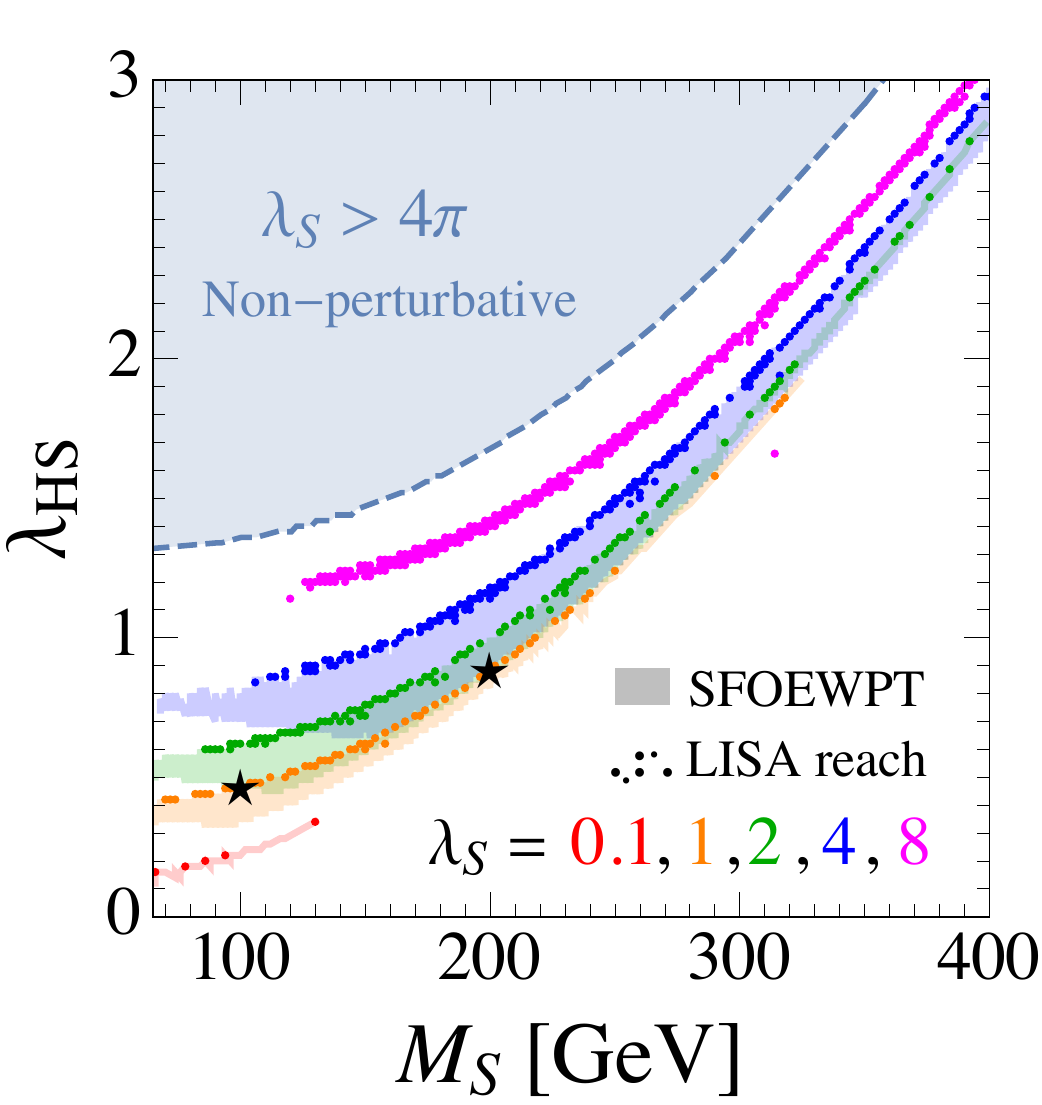}}\qquad
\subfigure{
\includegraphics[scale=0.35]{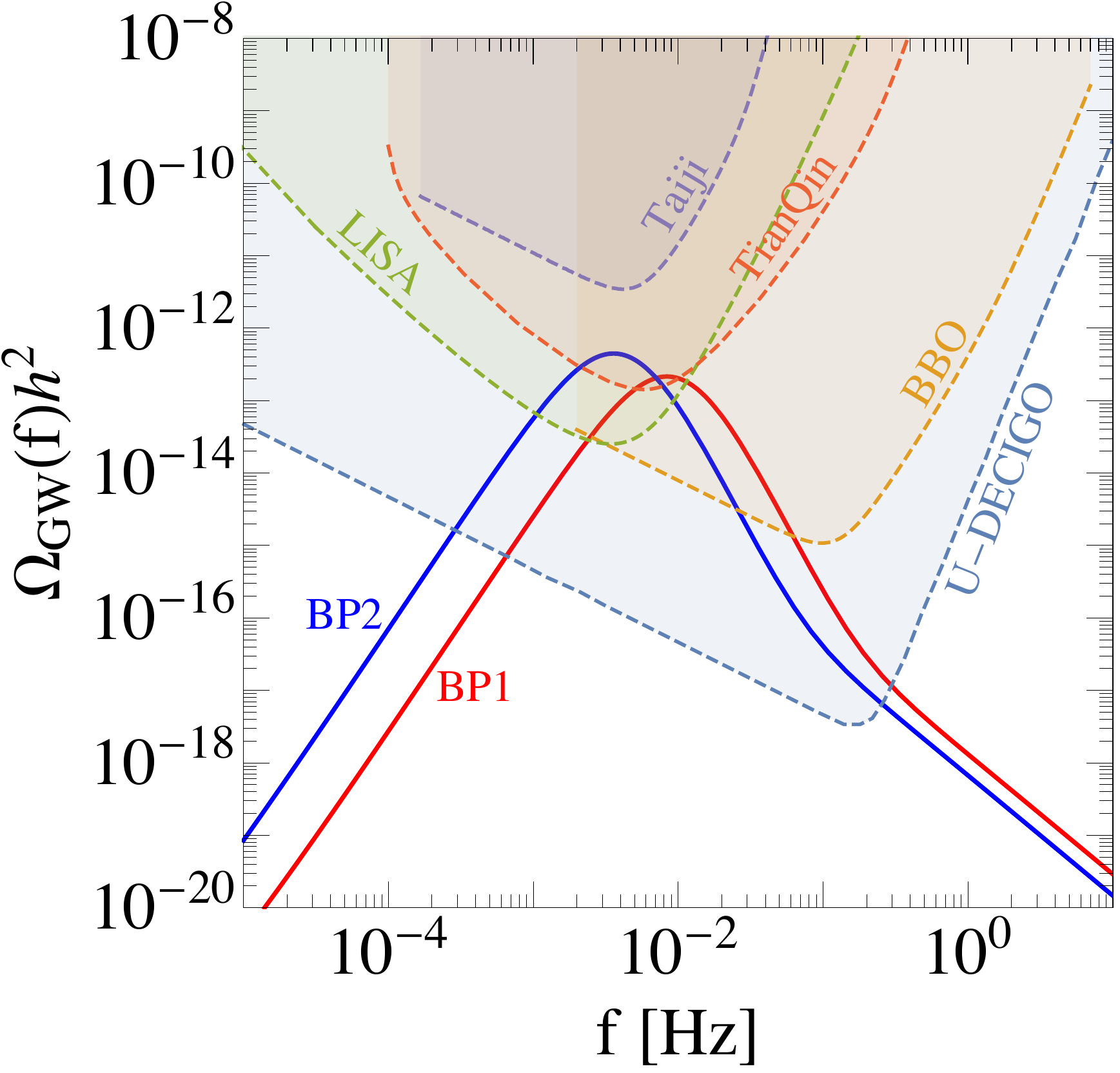}}
\caption{Left: parameter space allowed by a successful SFOEWPT. The colored shaded regions are for individual $\lambda_S$ values, except the light blue shaded area is for the non-perturbative region that $\lambda_S>4\pi$. The colored scatter points are the parameter regions that are detectable at the LISA. For $\lambda_S=8$, to avoid the plot to be too messy we omit the general SFOEWPT region and only show the LISA detectable points. The star-symbols highlight the two BPs in \Eq{BPs_1}. Right: the GW spectrum for the two BPs selected from the left panel.}
\label{fig:SFOEWPT}
\end{figure}

For a numerical study, we adopt the complete one-loop potential \Eq{VT} as the input. Using \Eq{tree-level_2}, we choose the three free parameter of the scalar potential as $M_S$, $\lambda_{HS}$ and $\lambda_S$. For a fixed $\lambda_S$, we scan over $M_S$ and $\lambda_{HS}$ and derive the SFOEWPT parameter space by solving \Eq{Linde} with the {\tt CosmoTransitions} package~\cite{Wainwright:2011kj} and checking \Eq{n_condition}. The results are shown as shaded regions in Fig.~\ref{fig:SFOEWPT}. We found that the shape of those regions match the analytical condition in \Eq{degenerate_V} very well. This is because in the two-step phase transition paradigm, the SFOEWPT is induced by the tree level potential barrier dominated by the $\lambda_{HS}h^2S^2/2$ term in the potential, thus a leading $T^2$ analysis already provides very good approximation. Due to the same reason, a sizable $\lambda_{HS}$ is needed for a successful SFOEWPT.  The correlation between SFOEWPT and the $\lambda_{HS}$ parameter provides an excellent channel for probing our mechanism at the collider, as will be shown in Section~\ref{sec:4tau}.

A SFOEWPT in the early universe may show detectable cosmological signals today, as the phase transition GWs typically peak around millihertz, within the sensitive region of a few next-generation space-based detectors, such as LISA~\cite{Audley:2017drz}, BBO~\cite{Crowder:2005nr}, TianQin~\cite{Luo:2015ght,Hu:2017yoc}, Taiji~\cite{Hu:2017mde,Guo:2018npi} and DECIGO~\cite{Kawamura:2011zz,Kawamura:2006up}. GWs are produced by three sources: bubble collision, sound waves and turbulence. For a SFOEWPT with a non-luminal terminal bubble velocity, the bubble collision contribution is negligible and the GWs are dominated by sound waves~\cite{Ellis:2018mja}. The GW spectrum today is described by
\be\label{GW}
\Omega_{\rm GW}(f)=\frac{1}{\rho_c}\frac{\rho_{\rm GW}}{d\ln f},
\ee
where $f$ is the frequency, $\rho_{\rm GW}$ is the GW energy density and $\rho_c$ is the critical energy density of the present universe. For a given SFOEWPT, \Eq{GW} can be expressed as numerical functions of the following three physical parameters~\cite{Grojean:2006bp,Caprini:2015zlo,Caprini:2019egz}:
\begin{enumerate}
\item $v_b$, the wall velocity of the expanding bubbles.
\item $\alpha$, the ratio of the SFOEWPT latent heat to the cosmic radiative energy density
\be
\alpha=\frac{1}{\rho_R(T_n)}\left(-\Delta V_T+T\frac{\partial \Delta V_T}{\partial T}\right)\Big|_{T_n},
\ee
where $\rho_R(T_n)=\pi^2g(T_n)T_n^4/30$ with $g(T_n)$ being the number of relativistic degrees of freedom, and $\Delta V_T$ is the free energy difference between the true and false vaca.

\item $\beta/H$, the inverse ratio of the time scale of the SFOEWPT and the universe expansion,
\be
\beta/H= T_n \frac{d}{dT}\left(\frac{S_3(T)}{T}\right)\Big|_{T_n}.
\ee
\end{enumerate}
In general, $\beta/H$ is related to the duration of the SFOEWPT hence the peak frequency of the GWs, while $\alpha$ is relevant to the signal strength.

We obtain the GWs spectrum using the formulae from the references mentioned above, with the energy budget from Ref.~\cite{Espinosa:2010hh}\footnote{See Ref.~\cite{Wang:2020nzm} for a recent study about the energy budget beyond the bag model equation of state.}. The bubble velocity is adopted as $v_b=0.6$ as a benchmark. Strictly speaking, the $\alpha$ and $\beta/H$ parameters should be calculated at the percolation temperature $T_p$~\cite{Megevand:2016lpr,Kobakhidze:2017mru,Ellis:2018mja,Ellis:2020awk,Wang:2020jrd}; however the supercooling effect in our scenario is not strong -- we have checked that $\alpha\lesssim1$ for our parameter space -- thus $T_p\approx T_n$ is a good approximation. We have taken into account the suppression factor $H\tau_{\rm sw}$ for the sound wave contribution due to the shortness of its duration compared to the Hubble time~\cite{Ellis:2018mja}\footnote{We adopt the suppression factor as $\Upsilon=1-(1+2H\tau_{\rm sw})^{-1/2}$~\cite{Guo:2020grp}.}. For the detectability of GWs signals we take the LISA as an example and evaluate the signal-to-noise-ratio (SNR)~\cite{Caprini:2015zlo}
\be
{\rm SNR}=\sqrt{\mT\int_{f_{\rm min}}^{f_{\rm max}} df\left(\frac{\Omega_{\rm GW}(f)}{\Omega_{\rm LISA}(f)}\right)^2},
\ee
with $\Omega_{\rm LISA}$ being the sensitive curve of the LISA detector~\cite{Caprini:2015zlo}, and $\mT=9.46 \times 10^7$ s the data-taking duration~\cite{Caprini:2019egz}. With ${\rm SNR}=10$ as the detectable threshold, we find that for a fixed $\lambda_S$, there is a narrow band in the $M_S-\lambda_{HS}$ plane that can be probed by the LISA, as shown in the left panel of Fig.~\ref{fig:SFOEWPT}. Fixing $\lambda_S=1$, we choose two benchmark points (BPs) as
\bea\label{BPs_1}
&&{\rm BP1}:\quad M_S=100~{\rm GeV},\quad \lambda_{HS}=0.46;\quad \alpha=0.13,\quad\beta/H=406,\nn\\ 
&&{\rm BP2}:\quad M_S=200~{\rm GeV},\quad\lambda_{HS}=0.88;\quad \alpha=0.05,\quad\beta/H=109,
\eea
and plot their GW spectra in the right panel of Fig.~\ref{fig:SFOEWPT} as an illustration.

\subsection{Transport equations and the generation of BAU}

Consider a SFOEWPT triggered by the vacuum decay $(0,w_n)\to(v_n,0)$ at $T_n$. Since EWBG happens in the vicinity of the bubble wall, it's convenient to work in the wall rest frame. Under the planar wall approximation, the scalars can be treated as background fields $\hat h$ and $\hat S$ that depend only on the spatial distance $z$ from the center of wall. Assuming $z\to\pm\infty$ is the broken/symmetric phase, the scalar backgrounds can be parametrized as
\be
\hat h(z)=\frac{v_n}{2}\left(1+\tanh\frac{z}{L_w}\right),\quad\hat S(z)=\frac{w_n}{2}\left(1-\tanh\frac{z}{L_w}\right),
\ee
with $L_w$ being the wall thickness. 
 
To realize another necessary condition for EWBG, i.e. the CP violation, we shall assume the singlet has the following leptophilic interactions with the SM fields via dimension-5 operators,
\be\label{dim-5}
\mL_5=\sum_{i,j}\frac{c_{ij}}{\Lambda}S\bar\ell_L^iHe_R^j+\hc,
\ee
where $i$, $j$ are generation indices, $c_{ij}$ are the (complex) Wilson coefficients, and $\Lambda$ is the cutoff scale of EFT. For simplicity, we assume the Willson coefficients are flavor-diagonal, i.e. $c_{ij}={\rm diag}\{c_e,c_\mu,c_\tau\}$. During a SFOEWPT, the effective mass of a lepton reads
\be\label{fermion_mass}
\bar m_i(z)=\frac{y_i}{\sqrt{2}}\hat h(z)+\frac{c_i}{\sqrt{2}\Lambda}\hat h(z)\hat S(z),
\ee
where $i=e$, $\mu$, $\tau$, and $y_i=\sqrt{2}M_i/v$ is the SM Yukawa coupling, with $M_i$ being the physical lepton mass at zero temperature. Since the effective mass is space-dependent, the complex phase of $c_i$ cannot be globally rotated away, resulting in physical CPV effects.

The Yukawa couplings for the first and second generation leptons are too small to generate a considerable CPV source~\cite{deVries:2018tgs}, thus we neglect them and focus on the $\tau$ lepton. In this case, the magnitude of $c_\tau$ can be absorbed into the definition of $\Lambda$, and we can rewrite the relevant part of \Eq{dim-5} as
\be\label{tau-5}
\mL_5\supset \frac{e^{i\phi_\tau}}{\Lambda_\tau}S\bar\ell_LH\tau_R+\hc,
\ee
such that $\phi_\tau$ is a pure CPV phase. Note that this operator is rather weakly constrained by the EDM experiments, especially in our scenario that $V_0$ is $\Z_2$ symmetric and $\ave{S}=0$ so that there is no tree level $h-S$ mixing at zero temperature.

In the plasma, the $\tau$ leptons participate in gauge, Yukawa, helicity-flipping and EW sphaleron interactions. We deal with those interactions following the standard two-step approach: in the first step we derive the generation and diffusion of the (lepton) chiral asymmetry by solving the Boltzmann equations of the ``fast processes'', and then in the second step the chiral asymmetry is converted into a baryon asymmetry via the EW sphaleron. Here EW sphaleron is the ``slow process'' and all other interactions are treated as ``fast''.

First we establish and solve the Boltzmann equations. Denoting the net particle density (i.e. the number density difference between particle and antiparticle) for left-handed third generation leptons, right-handed taus, and the Higgs bosons as $\ell=n_{\nu_{\tau L}}+n_{\tau_L}$, $\tau=n_{\tau_R}$ and $h=n_{G^+}+n_{H^0}$, respectively, the transport equations read~\cite{deVries:2018tgs}
\bea\label{Boltzmann}
v_w\ell'-D_\ell\ell''&=&\Gamma_M\left(\frac{\tau}{k_\tau}-\frac{\ell}{k_\ell}\right)+\Gamma_Y\left(\frac{\tau}{k_\tau}-\frac{\ell}{k_\ell}+\frac{h}{k_h}\right)-S_\tau\nn,\\
v_w\tau'-D_\tau\tau''&=&-\Gamma_M\left(\frac{\tau}{k_\tau}-\frac{\ell}{k_\ell}\right)-\Gamma_Y\left(\frac{\tau}{k_\tau}-\frac{\ell}{k_\ell}+\frac{h}{k_h}\right)+S_\tau,
\eea
where $\ell$, $\tau$ and $h$ are functions of the spatial coordinate $z$, and ``\,$\prime$\,'' represents the $d/dz$ operation. $v_w\in(0,1)$ is the bubble wall velocity with respect to the plasma {\it just in front of the wall}. Note that $v_w$ can differ from $v_b$ defined in Section~\ref{sec:GWs}, which is relative wall velocity to the plasma {\it at infinite distance}~\cite{No:2011fi}. $D_\ell=100/T_n$ and $D_\tau=380/T_n$ are respectively the diffusion constants for left- and right-handed leptons (note that they are much bigger than quark diffusion constant $D_q=6/T_n$)~\cite{Joyce:1994zn}. $\Gamma_M$ and $\Gamma_Y$ denote the helicity-flipping and Yukawa rates, respectively, and their detail definitions as well as the temperature-dependent coefficients $k_i$ are given in Appendix~\ref{app:thermal}. The EW gauge interactions are not written in \Eq{Boltzmann} because they are treated as in equilibrium, so that the two components of the same $SU(2)_L$ doublet share a common chemical potential. $S_\tau$ is the CPV source given by the closed time path method~\cite{Lee:2004we}
\be\label{CP_source}
S_\tau=\frac{v_w}{\pi^2}{\rm Im}\left[\bar m_\tau'm_\tau^*\right]J_\tau,
\ee
where $\bar m_\tau$ is the $z$-dependent effective mass in \Eq{fermion_mass}, while $J_\tau$ is a numerical factor whose expression is presented in Appendix~\ref{app:thermal}. \Eq{CP_source} clearly reveals that the number density difference between $\ell_L$ and $\bar\ell_L$ is sourced by the space-dependent imaginary part of the effective mass (namely the physical CPV phase) when the $\tau$ leptons are passing across the bubble wall.

There are three unknown functions $\ell$, $\tau$ and $h$ in \Eq{Boltzmann}, but only two equations (for leptons). To solve the equations self-consistently we need one more equation about $h$. However, the transport equation of $h$ involves the heavy quarks such as top and bottom, because they interact strongly with the Higgs. As a result, a complete treatment for lepton transport should also include the equations for quarks as well. Thanks to the small lepton Yukawa couplings, the impact of $h$ on lepton transport is negligible, hence we can reasonably approximate $h\approx0$ and avoid the complexity of the quark transport equations. In addition, the diffusion constant for left- and right-handed leptons can be approximated as the same, i.e. $D_\ell\approx D_\tau\approx100/T_n$, which provides a relation $\ell=-\tau$ and therefore \Eq{Boltzmann} is simplified into a single ordinary differential equation about $\ell$
\be\label{sim_Boltzmann}
-D_\ell\ell''+v_w\ell'+(\Gamma_M+\Gamma_Y)\left(\frac{1}{k_\ell}+\frac{1}{k_\tau}\right)\ell=-S_\tau,
\ee
which can be solved either semi-analytically or numerically. Ref.~\cite{deVries:2018tgs} has shown that \Eq{sim_Boltzmann} is a very good approximation to the complete treatment which takes into account $D_\ell\neq D_\tau$ and the heavy quark transport.

By numerically solving \Eq{sim_Boltzmann}, we get the non-zero chiral asymmetry $\ell$. This asymmetry is then converted into a baryon-to-entropy ratio~\cite{deVries:2018tgs}
\be\label{nB}
\eta_B=\frac{n_B}{s}=-\frac{3\Gamma_{\rm ws}}{2sD_q\alpha_+}\int_{-\infty}^0dz\ell(z)\exp\{-\alpha_-z\},
\ee
where $D_q=6/T_n$ is the quark diffusion constant~\cite{Joyce:1994zn}, $\Gamma_{\rm ws}=6\kappa\alpha_W^5T_n$ is the EW sphaleron rate with $\kappa\approx18$~\cite{DOnofrio:2014rug}, and
\be
\alpha_\pm=\frac{v_w\pm\sqrt{15D_q\Gamma_{\rm ws}+v_w^2}}{2D_q}.
\ee
The integration region of \Eq{nB} is limited within the EW symmetric phase where the EW sphaleron is active. The generated baryon number survives until today, becoming the observed BAU. Note that the dimension-5 operator in \Eq{tau-5} is odd under the $\Z_2$ transformation, so does the CPV source $S_\tau$ in \Eq{CP_source}. Consequently, if there were an exact $\Z_2$ for $S$ in the potential, the $+S$ and $-S$ transitions equally distributed in different patches of the universe finally give a zero net BAU on average. This consideration gives another motivation for the small $\Z_2$-breaking term~\cite{Espinosa:2011eu}.

\begin{figure}
\centering
\subfigure{
\includegraphics[scale=0.415]{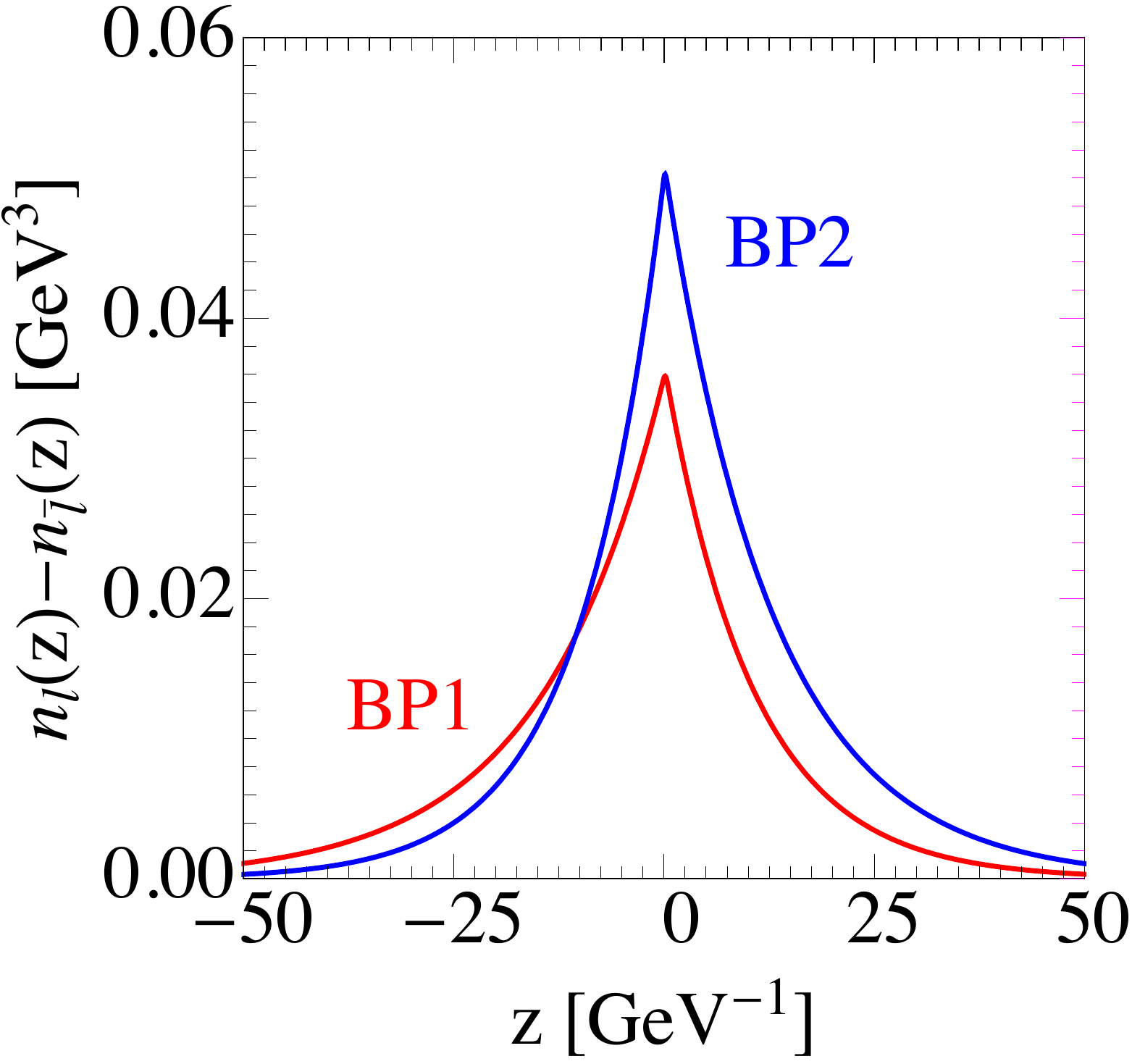}}\qquad
\subfigure{
\includegraphics[scale=0.4]{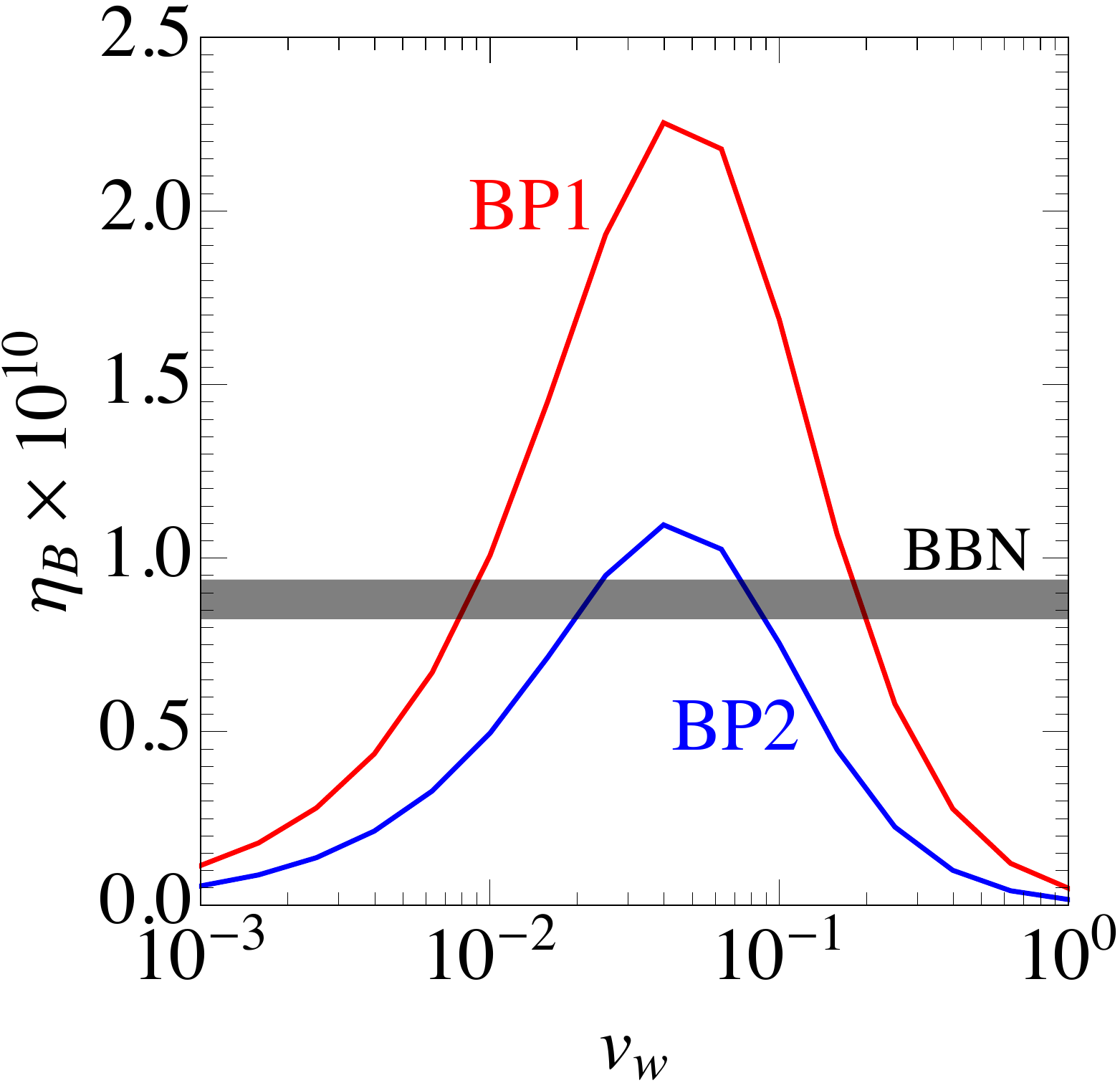}}
\caption{Left: the chiral asymmetry $\ell(z)\equiv n_{\ell}(z)-n_{\bar\ell}(z)$ solved from the transport equation (\ref{sim_Boltzmann}). Right: the BAU as a function of wall velocity $v_w$. The gray band represents the BBN fitted result of $\eta_B$. The CPV phase is $\phi_\tau=0.02$ for both panels.}
\label{fig:EWBG}
\end{figure}

For illustration, we consider the two SFOEWPT BPs in \Eq{BPs_1} of Section~\ref{sec:GWs}. Their SFOEWPT profiles are
\bea\label{BPs}
&&{\rm BP1}:\quad T_n=64.44~{\rm GeV},\quad v_n=239.06~{\rm GeV},\quad w_n=132.13~{\rm GeV};\nn\\
&&{\rm BP2}:\quad T_n=94.32~{\rm GeV},\quad v_n=220.55~{\rm GeV},\quad w_n=108.97~{\rm GeV},
\eea
and we take the wall thickness $L_w=15/T_n$ according to Ref.~\cite{Konstandin:2014zta}. The two BPs are fed into the Boltzmann equation (\ref{sim_Boltzmann}) to get the chiral asymmetry and then we use \Eq{nB} to evaluate the BAU. The CPV phase and cutoff scale are chosen to be $\phi_\tau=0.02$ and $\Lambda_\tau=10$ TeV,\footnote{The choice of $\Lambda_\tau=10$ TeV seems much larger than the scale adopted in Ref.~\cite{deVries:2018tgs}, which studies the CPV dimension-6 operators in SM EFT and uses $\Lambda_\tau=1$ TeV. However, the normalization schemes are different for these two researches: in Ref.~\cite{deVries:2018tgs}, the CPV operator is $(iy_\tau/\Lambda_\tau^2)|H|^2\bar\ell_LH\tau_R$, which has an additional suppression factor $y_\tau\approx0.01$; while in this article, the relevant operator is $(e^{i\phi_\tau}/\Lambda_\tau)S\bar\ell_LH\tau_R$. Once taken into account this difference, the amounts of CPV in our scheme and Ref.~\cite{deVries:2018tgs} are of the same order (or more specifically, for the selected benchmarks, the CPV effects in our article are several times larger).} respectively. The chiral asymmetry $\ell(z)\equiv n_\ell(z)-n_{\bar\ell}(z)$ is shown in the left panel of Fig.~\ref{fig:EWBG}, where we can recognize that the asymmetry is generated on the bubble wall, and then diffuses to both the EW symmetric and broken phases. Only the chiral asymmetry in the symmetric phase is responsible for generating BAU, and we plot the resultant BAU as a function of $v_w$ in the right panel of Fig.~\ref{fig:EWBG}. The observed BAU is shown in gray band and we find it can be achieved for $v_w\sim0.01$ or 0.1 in the figure, where $\phi_\tau=0.02$ is fixed. Since $\phi_\tau$ is actually a free parameter, we conclude that our mechanism can explain the BAU within a vast parameter space of $\phi_\tau$ and $v_w$.

\section{Collider search: the $4\tau$ final state}\label{sec:4tau}

Complementary to the space-based GW detectors, the terrestrial collider experiments also serve as a powerful probe for the SFOEWPT by either measuring the Higgs potential shape or directly detecting the relevant BSM physics around TeV scale~\cite{Profumo:2014opa,Cao:2017oez,Alves:2018oct,Zhou:2020idp,Chen:2019ebq,Huang:2016cjm,Kozaczuk:2019pet,Papaefstathiou:2020iag,Alves:2020bpi}. In this section we assess the possibility of detecting the lepton-mediated EWBG at the 14 TeV LHC and the 27 TeV HE-LHC~\cite{Abada:2019ono}. At the collider, the real singlet in the xSM can be probed by the resonant or non-resonant di-Higgs production, the modified Higgs-fermion or Higgs-gauge couplings, etc. However, most of the proposed channels require the mixing between $S$ and $h$. If unfortunately the $S-h$ mixing is negligible, e.g. in the scenario considered in Section~\ref{sec:GWs} that $S$ has an approximate $\Z_2$ symmetry in the potential and $\ave{S}|_{T=0}=0$, then it would be very challenging to test the SFOEWPT at the collider~\cite{Ashoorioon:2009nf,Curtin:2014jma}. However, in our scenario, there exists the following process at a $pp$ collider,
\be\label{4tau}
gg\to h^*\to SS\to\tau^+\tau^-\tau^+\tau^-,
\ee
i.e. the pair production of $S$ through an off-shell Higgs (mediated by the $\lambda_{HS}v\,hS^2\subset\lambda_{HS}|H|^2S^2$ vertex), followed by the $S\to\tau^+\tau^-$ decay (mediated by the dimension-5 operator $(c_\tau/\Lambda_\tau)S\bar\ell_LH\tau_R$). This robust channel is present even in the absence of a $h-S$ mixing. Since the $\lambda_{HS}$ coupling is required by a SFOEWPT, while the dimension-5 operator is required by the CPV, both two necessary ingredients of our model is tested by the single process in \Eq{4tau}.

\begin{figure}
\centering
\subfigure{
\includegraphics[scale=0.55]{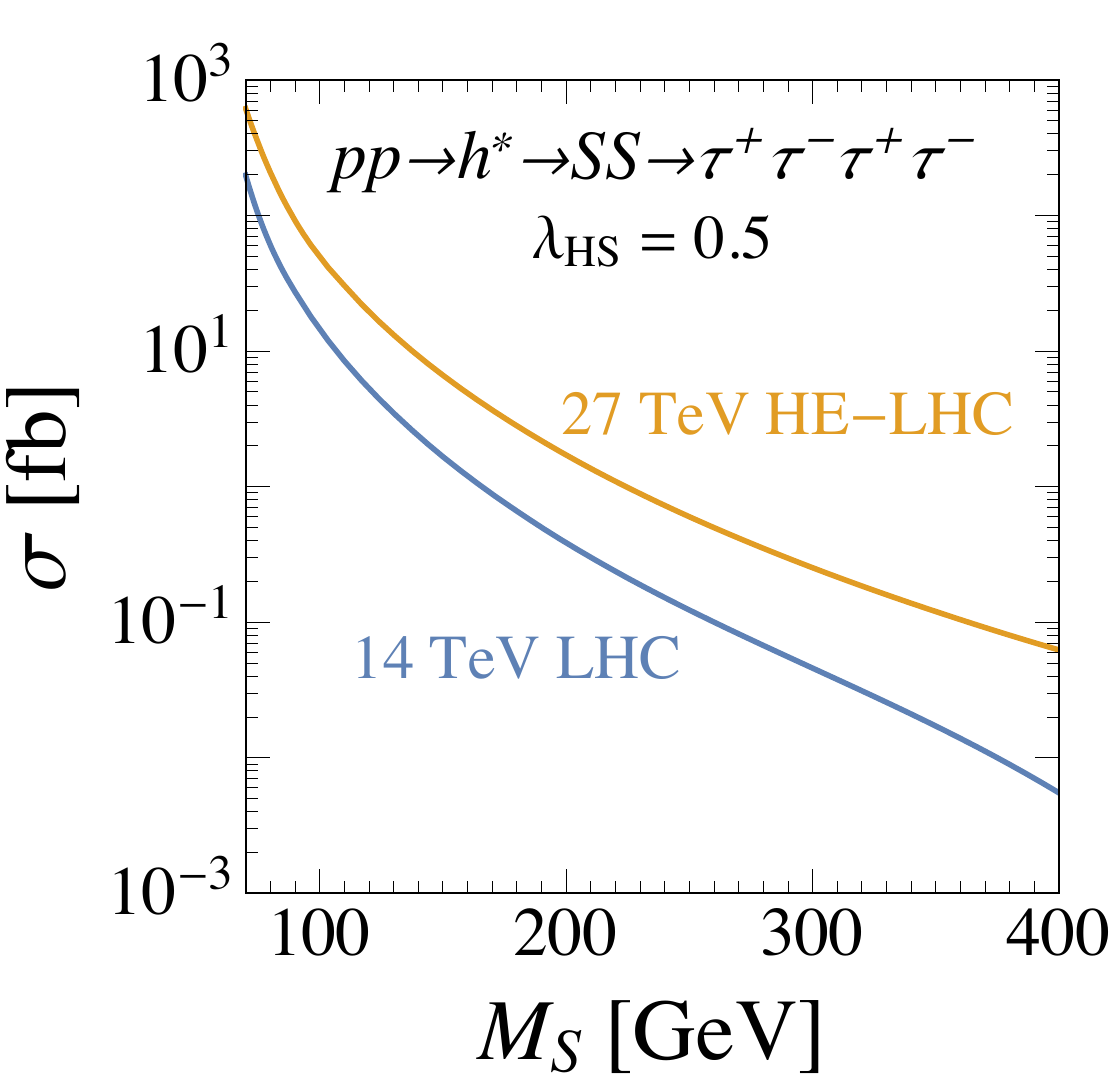}}
\caption{Production rates of $pp\to h^*\to SS\to\tau^+\tau^-\tau^+\tau^-$ at the 14 TeV LHC and 27 TeV HE-LHC. We have assumed $\lambda_{HS}=0.5$ and $\Br(S\to\tau^+\tau^-)=100\%$.}
\label{fig:SS}
\end{figure}

The cross sections of \Eq{4tau} are shown as functions of $M_S$ in Fig.~\ref{fig:SS}, where we have set $\lambda_{HS}=0.5$ and assumed $S$ decays exclusively to $\tau^+\tau^-$. Note that the production rate is proportional to $\lambda_{HS}^2$, thus this channel is sensitive to the SFOEWPT parameter space, which requires a sizable $\lambda_{HS}$ (see the left panel of Fig.~\ref{fig:SFOEWPT}). The decay branching ratio for a $\tau$ lepton is $\sim35\%$ and $\sim65\%$ for leptonic and hadronic channels, respectively. The leptonic decay yields a charged lepton ($e^\pm$ or $\mu^\pm$) and missing transverse momentum, while the hadronic decay typically results in a $\tau$-jet~\cite{Bagliesi:2007qx}. Therefore, the $4\tau$ final state can be categorized into the following final states directly available at the detector:
\be
4\tau_j~(17.9\%),\quad 1\ell3\tau_j~(38.4\%),\quad 2\ell2\tau_j~(31.1\%),\quad 3\ell1\tau_j~(11.1\%),\quad 4\ell~(1.5\%),
\ee
where $\ell$ denotes the charged leptons $e^\pm$ and $\mu^\pm$, $\tau_j$ denotes the $\tau$-jets, and the numbers in the brackets are the corresponding probabilities. We find that the $1\ell3\tau_j$ and $2\ell2\tau_j$ channels are hopefully to be probed, as they enjoy considerable cross section fractions and have charged leptons to trigger the events. While the $2\ell2\tau_j$ final state is searched at the LHC in supersymmetry-relevant experiments~\cite{Aaboud:2018zeb}, the $1\ell3\tau_j$ final state has not yet been dedicatedly searched at the LHC.

\begin{table}
\scriptsize\renewcommand\arraystretch{1.5}\centering
\begin{tabular}{|c|c|c|c|c|c|c|c|c|c|}\hline
Unit: fb & \tabincell{c}{Signal\\ BP1} & \tabincell{c}{Signal\\ BP2} & $W^\pm+{\rm jets}$ & $Z+{\rm jets}$ & $t\bar t$ & $W^\pm\tau^+\tau^-j$ & $\tau^+\tau^-+{\rm jets}$ & $\tau^+\tau^-\tau^+\tau^-$ \\ \hline\hline
\multicolumn{9}{|c|}{14 TeV LHC}\\ \hline
 Before & $12.3$ & $1.19$ & $1.45\times10^6$ & $6.18\times10^5$ & $1.21\times10^5$ & $129$ & $1.49\times10^5$ & $7.15$ \\ \hline
 Cut I & $1.76$ & $0.352$ & $2.43\times 10^5$ & $5.91\times10^4$ & $6.73\times10^4$ & $34.5$ & $6.35\times10^3$ & $0.511$ \\ \hline
 Cut II & $0.0733$ & $0.0269$ & $0.832$ & $3.28$ & $3.41$ & $0.152$ & $0.841$ & $0.0378$ \\ \hline
 Cut III & $0.0661$ & $0.0245$ & $0.681$ & $2.64$ & $0.243$ & $0.134$ & $0.762$ & $0.0356$ \\ \hline\hline
 \multicolumn{9}{|c|}{27 TeV HE-LHC}\\ \hline
 Before & $42.7$ & $5.30$ & $4.10\times10^6$ & $1.59\times10^6$ & $1.06\times10^6$ & $321$ & $3.34\times10^5$ & $13.4$ \\ \hline
 Cut I & $6.74$ & $1.64$ & $6.66\times10^5$ & $1.69\times10^5$ & $5.55\times10^5$ & $95.8$ & $1.72\times10^4$ & $1.19$ \\ \hline
 Cut II & $0.267$ & $0.115$ & $2.54$ & $13.9$ & $45.7$ & $0.369$ & $2.23$ & $0.0724$ \\ \hline
 Cut III & $0.245$ & $0.103$ & $2.05$ & $10.9$ & $9.14$ & $0.315$ & $1.87$ & $0.0635$ \\ \hline
\end{tabular}
\caption{Cut flows for backgrounds and the two signal BPs at the 14 TeV LHC and 27 TeV HE-LHC. The two signal BPs are from \Eq{BPs_1} in Section~\ref{sec:GWs}.}\label{tab:cut_flows}
\end{table}

In this work, we focus on the $1\ell3\tau_j$ channel, whose main backgrounds are the SM $W^\pm+{\rm jets}$, $Z+{\rm jets}$, $t\bar t$, $W^\pm\tau^+\tau^- j$, $\tau^+\tau^-+{\rm jets}$ and $\tau^+\tau^-\tau^+\tau^-$ processes. We perform a collider simulation using the packages {\tt FeynRules}~\cite{Alloul:2013bka} (to write the UFO model file~\cite{Degrande:2011ua} for our scenario), {\tt MadGraph5\_aMC@NLO}~\cite{Alwall:2014hca} (to generate parton level events for signal and backgrounds) and {\tt Pythia8}~\cite{Sjostrand:2007gs} and {\tt Delphes}~\cite{deFavereau:2013fsa} (for parton shower and fast detector simulation, respectively). The $W^\pm+{\rm jets}$, $Z+{\rm jets}$ and $\tau^+\tau^-+{\rm jets}$ backgrounds are realized by matching $W^\pm jjj$ to $W^\pm jj$ process, matching $Zjj$ to $Zj$ process, matching $\tau^+\tau^-jj$ to $\tau^+\tau^-j$ process, respectively. The $t\bar t$ is also matched to $+1~{\rm jet}$ final state. The inclusive decay of $\tau$ lepton is implemented by the {\tt Pythia8} package.

To suppress the backgrounds, we apply the following cuts to the events:
\begin{enumerate}
\item Exactly 1 charged lepton with $p_T^\ell>25$ GeV and $|\eta_\ell|<2.5$, and at least 3 jets with $p_T^j>20$ GeV and $|\eta_j|<5$.
\item At least 3 $\tau$-tagged jets. The tagging and mistag rates for a $\tau$-jet are set to be 60\% and 1\%, respectively.
\item Veto any event with $b$-tagged jets. The $b$-tagging efficiency is chosen as 77\%, and the mistag rate for $c$ and other light quarks are 17\% and 0.75\%, respectively~\cite{Aaboud:2018xpj}.
\end{enumerate}
The cut flows for the backgrounds and the two signal BPs are listed in Table~\ref{tab:cut_flows}. We can see that the requirement of three $\tau$-jets significantly suppresses the backgrounds, while the $b$-veto cut works very well in reducing the $t\bar t$ background.

\begin{figure}
\centering
\subfigure{
\includegraphics[scale=0.55]{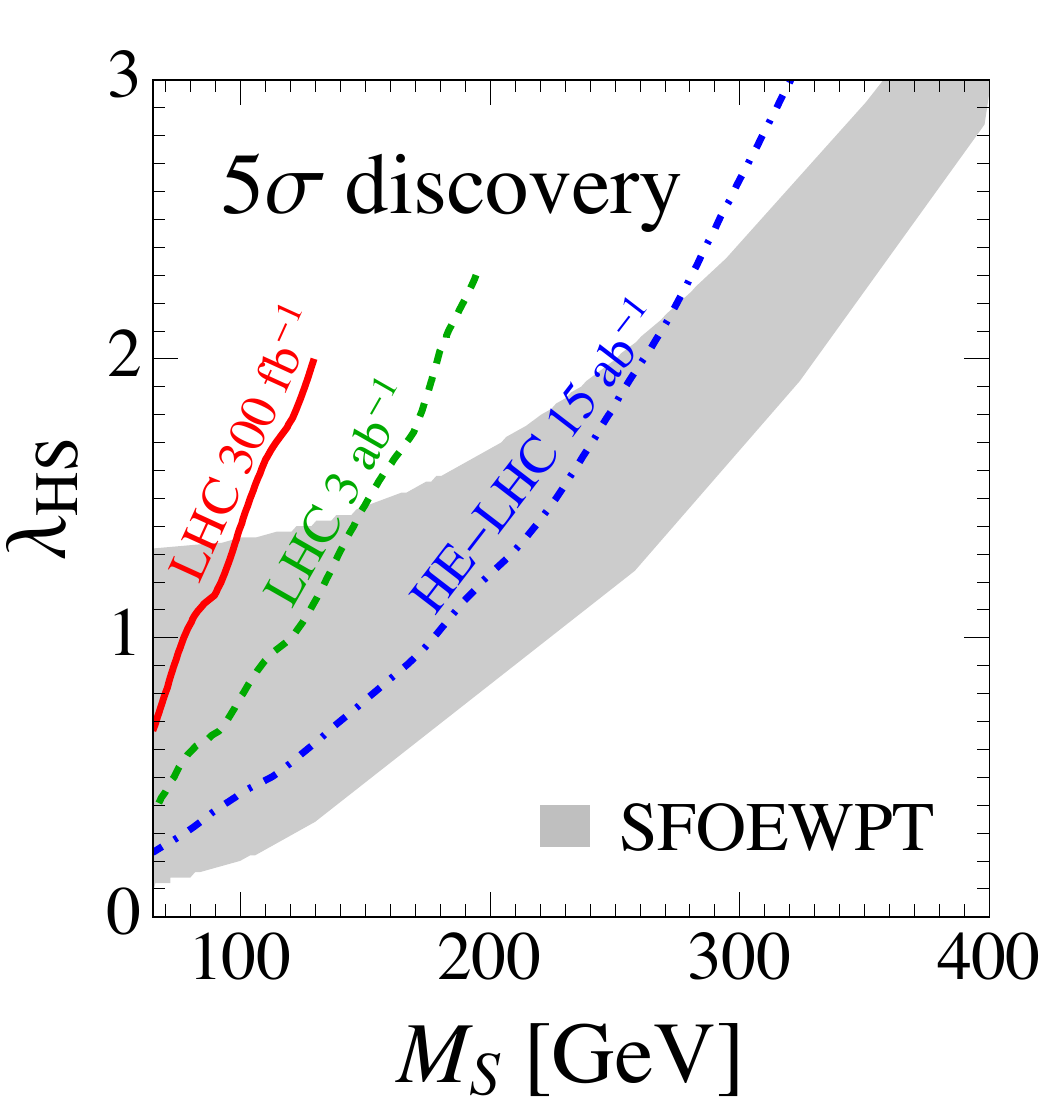}}\qquad
\subfigure{
\includegraphics[scale=0.55]{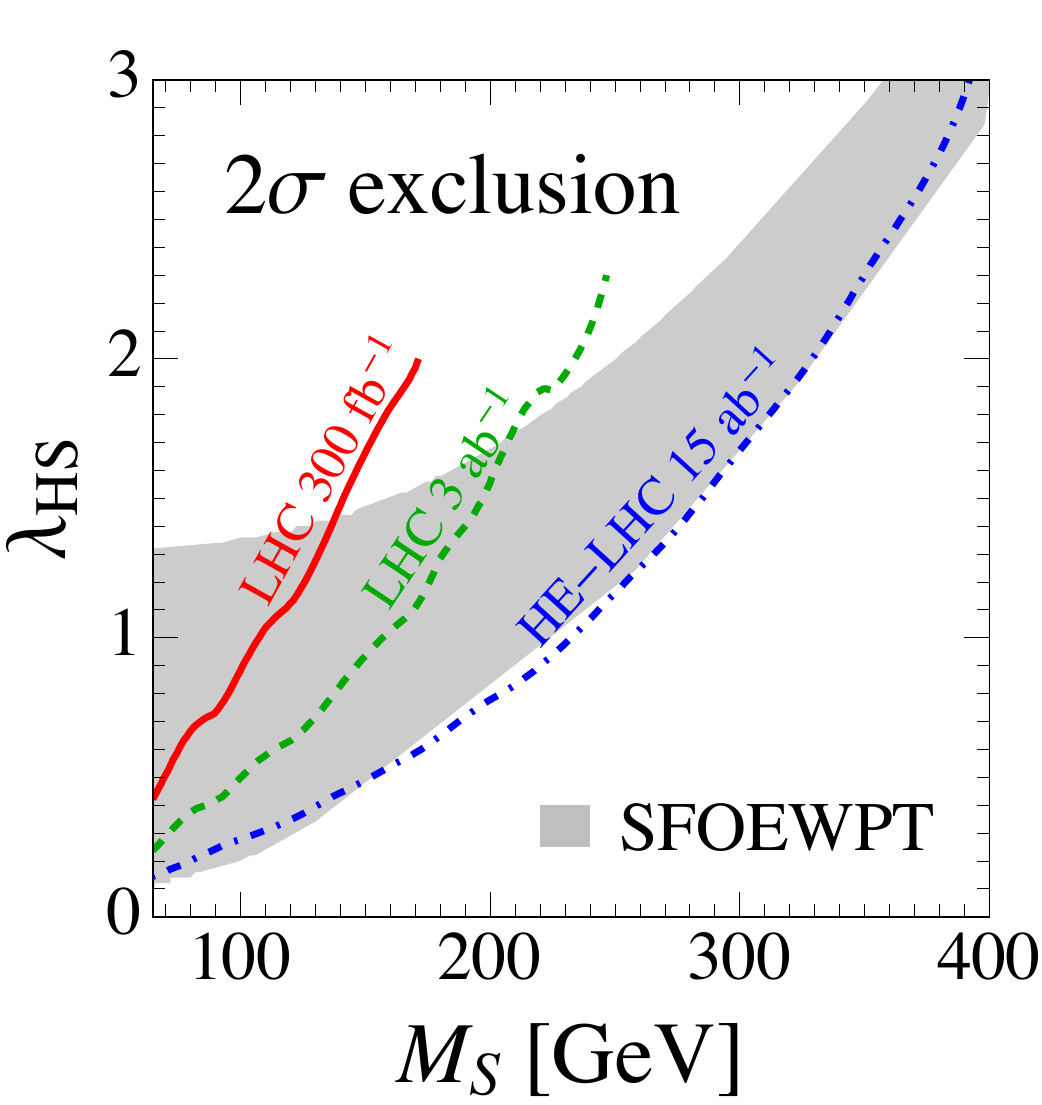}}
\caption{Probing the SFOEWPT parameter space at the LHC and HE-LHC. The left and right panels show the $5\sigma$ discovery and $2\sigma$ exclusion limits, respectively. The gray shaded region labeled as ``SFOEWPT'' is derived by varying $\lambda_S\in(0,4\pi)$.}
\label{fig:collider}
\end{figure}

Given the cut flow data, now we can derive the signal significance for a given set of $(M_S,\lambda_{HS})$ and integrated luminosity $\mL$:
\be
{\rm Significance}=\frac{\sigma_S\epsilon_S}{\sqrt{\sigma_B\epsilon_B}}\sqrt{\mL},
\ee
where $\sigma_{S,B}$ and $\epsilon_{S,B}$ denote the cross sections and cut efficiencies for signal and total background, respectively. We define the significance equal to $5\sigma$ and $2\sigma$ as the experimental discovery and exclusion limits of the collider at a given integrated luminosity, respectively. Note that the collider phenomenology is irrelevant to the singlet quartic coupling $\lambda_S$, while the SFOEWPT parameter space depends on $\lambda_S$, see the left panel of Fig.~\ref{fig:SFOEWPT}. In Fig.~\ref{fig:collider}, we obtain the whole SFOEWPT parameter space at the $\lambda_{HS}-M_S$ plane by varying $\lambda_S$ from 0 to the non-perturbative limit $4\pi$, and taking the envelop of the SFOEWPT regions. The expected reach of different collider setups is shown in the figure with solid or dashed lines with different colors. We can see that those reach curves have similar shapes with the SFOEWPT parameter space (gray shaded region), because both the $gg\to h^*\to SS$ process and the SFOEWPT prefer a large $\lambda_{HS}$. A considerable fraction of the SFOEWPT space can be probed at the HL-LHC (namely the LHC with an integrated luminosity of 3 ab$^{-1}$), while almost all the SFOEWPT parameter space can be covered by the 27 HE-LHC with an integrated luminosity of 15 ab$^{-1}$.

Note that the results in Fig.~\ref{fig:collider} are for the $1\ell3\tau_j$ channel only. We have also checked that the $2\ell2\tau_j$ channel shows a similar reach by using the backgrounds simulated in Ref.~\cite{Aaboud:2018zeb}. Therefore, the combination of those two channels can provide an even better probe for our mechanism. Finally, we emphasize that the collider phenomenology presented in this section is mainly determined by the existence of the dimension-5 operator $(c_\tau/\Lambda_\tau)S\bar\ell_LH\tau_R$, but not sensitive to its CPV phase. If in the future we really detected an excess in the $4\tau$ channel, then a further analysis should be performed to reveal its CP structure.

\section{Conclusion}\label{sec:conclusion}

In this article, we have proposed a novel EWBG mechanism mediated by $\tau$ lepton transport. The SM is extended with a real singlet to trigger the SFOEWPT and with a set of dimension-5 operators to get sufficient CPV source. For simplicity, we assume a scalar potential that is symmetric under $S\to-S$ at tree level, and demonstrate that it is able to realize the SFOEWPT with the real scalar mass at $\mO(100~{\rm GeV})$. By establishing and solving the Boltzmann equations, we have shown that this model is able to account for current BAU using the $\tau$-mediated EWBG.

Our model is testable in current or near-future experiments. On one hand, the phase transition GWs can be probed by the near-future space-based detectors (e.g. LISA); on the other hand, the pair production of singlets via the off-shell Higgs leads to distinguishable $4\tau$ final state at hadron colliders such as LHC and HE-LHC. This $pp\to h^*\to SS\to\tau^+\tau^-\tau^+\tau^-$ process is a characteristic channel of our mechanism, as the production of $S$ pair is induced by the $\lambda_{HS}$ term which is required by the SFOEWPT, while the decay of $S$ is induced by the dimension-5 operator which is needed for CPV. A detailed collider simulation shows that the EWBG parameter space can be efficiently explored through the $1\ell3\tau_j$ channel.

\acknowledgments

I am grateful to Ligong Bian, Peisi Huang, Benoit Laurent, Zhen Liu and Yehonatan Viernik for the useful discussions. I also thank Huai-Ke Guo for the discussion about the sound wave period of GWs. This work is supported by the Grant Korea NRF-2019R1C1C1010050.

\appendix
\section{The complete one-loop scalar potential}\label{app:one-loop}

The one-loop Coleman Weinberg potential in $\overline{\rm MS}$ scheme is
\be
V_{\rm CW}=\sum_j n_j\frac{m_j^4}{64\pi^2}\left(\ln\frac{m_j^2}{\mu^2}-C_j\right),
\ee
where the subscript $j$ runs over all particles of the model, and he field-dependent masses are
\bea
&&m_h^2=\mu_H^2+3\lambda_Hh^2+\lambda_{HS}S^2,\quad
m_S^2=\mu_S^2+3\lambda_SS^2+\lambda_{HS}h^2,\quad
m_{hS}^2=2\lambda_{HS}hS,\nn\\
&&m_{G^{\pm,0}}^2=\mu_H^2+\lambda_H h^2+\lambda_{HS}S^2,
\eea
for the scalars (where $G^{\pm,0}$ denote the Goldstone modes of the Higgs doublet) and
\be
m_W^2=\frac{g^2}{4}h^2,\quad m_Z^2=\frac{g^2+g'^2}{4}h^2,\quad m_t^2=\frac{y_t^2}{2}h^2.
\ee
for the gauge bosons and the top quark. The numerical factors are
\be\label{n}
\begin{cases}~n_j=1,\quad C_j=3/2,\quad&\text{for scalar fields;}\\
~n_j=3,\quad C_j=5/6,\quad&\text{for vector fields;}\\
~n_j=-4N_c,\quad C_j=3/2,\quad&\text{for Dirac fermions,}
\end{cases}
\ee
where $N_c$ is the color number, which is 3 for a quark and 1 for a lepton. For our numerical study, the renormalization scale is chosen as the top mass, i.e. $\mu=M_t=173.2$ GeV.

The counter term is defined as
\be\label{VCT}
V_{\rm CT}=\frac{\delta\mu_H^2}{2}h^2+\frac{\delta\mu_S^2}{2}S^2+\frac{\delta\lambda_H}{4}h^4+\frac{\delta\lambda_S}{4}S^4+\frac{\delta\lambda_{HS}}{2}h^2S^2.
\ee
We choose the following renormalization conditions,
\bea
&&(V_{\rm CW}+V_{\rm CT})\big|_{(0,w)}=0,\quad \frac{\partial (V_{\rm CW}+V_{\rm CT})}{\partial h}\Big|_{(v,0)}=\frac{\partial (V_{\rm CW}+V_{\rm CT})}{\partial S}\Big|_{(0,w)}=0,\nn\\
&&\frac{\partial^2 (V_{\rm CW}+V_{\rm CT})}{\partial h^2}\Big|_{(v,0)}=\frac{\partial^2 (V_{\rm CW}+V_{\rm CT})}{\partial S^2}\Big|_{(v,0)}=0,
\eea
to determine the coefficients in \Eq{VCT}, such that the tree-level relations in \Eq{tree-level_1} and \Eq{tree-level_2} still hold at one-loop level.

The thermal integral correction is
\be
\tilde V_T=\sum_jn_j\frac{T^4}{2\pi^2}J_j\left(\frac{m_j^2}{T^2}\right),
\ee
where $j$ again runs over all particles and $m_j$ again is the field-dependent mass given in last subsection. The thermal integral are
\be
J_{B/F}(y)=\int_0^\infty x^2dx\ln\left(1\mp e^{-\sqrt{x^2+y}}\right),
\ee
where $B/F$ is for bosons/fermions. Finally, the daisy resummation term is
\be
V_{\rm daisy}=-\sum_{j'}\frac{T}{12\pi}\left[\left(m_{j'}^{T}\right)^3-m_{j'}^3\right],
\ee
where $j'$ runs over the scalars and longitudinal modes of the vector bosons, and $m_{j'}^T$ is the Debye mass whose expression can be found in Refs.~\cite{Carrington:1991hz,Espinosa:2011ax}.

For $m_j^2/T^2\lesssim2$, the thermal integrals can be approximated as
\be
\frac{T^4}{2\pi^2}J_B\left(\frac{m_j^2}{T^2}\right)\approx\frac{T^2}{24}m_j^2,\quad \frac{T^4}{2\pi^2}J_F\left(\frac{m_j^2}{T^2}\right)\approx-\frac{T^2}{48}m_j^2.
\ee
Using this expansion, and ignoring the Coleman-Weinberg potential, the counter terms and the daisy resummation, we can get the leading $T^2$ approximation for $V_T$, i.e. \Eq{VT_sim}.

\section{Coefficients in the Boltzmann equations}\label{app:thermal}

In this appendix we list the definitions of the coefficients exist in the transport equation \Eq{Boltzmann}. The $k_i$ coefficients are given by~\cite{deVries:2017ncy,Fuchs:2020pun}
\be
k_i=\tilde k_i\frac{c_{B/F}}{\pi^2}\int_{a_i}^\infty dx\frac{xe^x}{(e^x\mp1)^2}\sqrt{x^2-a_i^2},
\ee
where the upper/lower sign is for bosons/fermions, $c_{B/F}=3$ or 6 for bosons/fermions, $\tilde k_i$ is the physical degrees of freedom in a multiplet ($\tilde k_h=4$, $\tilde k_\ell=2$ and $\tilde k_\tau=1$ in our scenario), and $a_i=\sqrt{{\rm Re}[\delta m_i^2(T_n)]}/T_n$, with the real part of the thermal masses being~\cite{Enqvist:1997ff}
\bea\label{thermal_mass}
{\rm Re}[\delta m_h^2(T)]&=&\left[\frac{3}{16}g^2+\frac{1}{16}g'^2+\frac{1}{12}\sum_j\left(y_{e^j}^2+3y_{u^j}^2+3y_{d^j}^2\right)\right]T^2,\nn\\
{\rm Re}[\delta m_\ell^2(T)]&=&\left(\frac{3}{32}g^2+\frac{1}{32}g'^2+\frac{1}{16}y_\tau^2\right)T^2,\nn\\
{\rm Re}[\delta m_\tau^2(T)]&=&\left(\frac{1}{8}g'^2+\frac18y_\tau^2\right)T^2.
\eea

The interactions rates in \Eq{Boltzmann} are~\cite{Fuchs:2020pun}
\begin{multline}
\Gamma_M=\frac{3}{\pi^2T_n^3}|\bar m_\tau|^2\int_0^\infty\frac{k^2dk}{\omega_L\omega_R}\times\\
{\rm Im}\left[\frac{h_F(\epsilon_R)+h_F(\epsilon_L)}{\epsilon_R+\epsilon_L}\left(\epsilon_L\epsilon_R+k^2\right)-\frac{h_F(\epsilon_R^*)+h_F(\epsilon_L)}{\epsilon_R^*-\epsilon_L}\left(\epsilon_L\epsilon_R^*-k^2\right)\right],
\end{multline}
and
\bea\label{new}
\Gamma_Y&=&\frac{3y_\tau^2}{4\pi^3T_n^2}(m_h^2-m_\ell^2-m_\tau^2)\int_{m_\tau}^\infty d\omega_Rh_F(\omega_R)
\ln\left(\frac{e^{\omega_R/T_n}+e^{\omega_-/T_n}}{e^{\omega_R/T_n}+e^{\omega_+/T_n}}\frac{e^{\omega_+/T_n}-1}{e^{\omega_-/T_n}-1}\right)\nn\\
&&+\frac{3\zeta_3}{32\pi^3}g^2y_\tau^2T_n\ln\left(\frac{8T_n^2}{m_\ell^2}\right),
\eea
where $\zeta_3\approx1.202$, $\bar m_\tau$ is the effective mass in \Eq{fermion_mass}, $m_{\ell,\tau,h}^2$ are short for the thermal masses in \Eq{thermal_mass}, and
\begin{multline}
\omega_\pm=\frac{1}{2m_\tau^2}\Big[\omega_R|m_h^2+m_\tau^2-m_\ell^2|\\
\pm\sqrt{(\omega_R^2-m_\tau^2)(m_\tau^2-(m_\ell+m_h)^2)(m_\tau^2-(m_\ell-m_h)^2)}\Big].
\end{multline}
The $\omega_{R/L}$ and $\epsilon_{R/L}$ functions are
\be
\omega_{R/L}(\textbf{k})=\sqrt{|\textbf{k}|^2+{\rm Re}[\delta m_{\tau/\ell}^2(T)]},\quad \epsilon_{R/L}(\textbf{k})=\omega_{R/L}(\textbf{k})-i\Gamma_\tau,
\ee
where $\Gamma_\tau\approx0.002\,T_n$~\cite{Elmfors:1998hh}. The $n_F$ and $h_F$ functions are
\be
n_F(k_0)=\frac{1}{e^{k_0/T_n}+1},\quad h_F(k_0)=\frac{e^{k_0/T_n}}{(e^{k_0/T_n}+1)^2}.
\ee

Finally, the factor $J_\tau$ in the CPV source of \Eq{CP_source} is given by~\cite{Lee:2004we,Cirigliano:2006wh,Fuchs:2020pun}
\be
J_\tau=\int_0^\infty\frac{k^2dk}{\omega_L\omega_R}{\rm Im}\left[\frac{n_F(\epsilon_L)-n_F(\epsilon_R^*)}{(\epsilon_L-\epsilon_R^*)^2}(\epsilon_L\epsilon_R^*-k^2)+\frac{n_F(\epsilon_L)+n_F(\epsilon_R)}{(\epsilon_L+\epsilon_R)^2}(\epsilon_L\epsilon_R+k^2)\right].
\ee

\bibliographystyle{JHEP-2-2.bst}
\bibliography{references}

\providecommand{\href}[2]{#2}\begingroup\raggedright\begin{thebibliography}{100}

\bibitem{Tanabashi:2018oca}
{\scshape Particle Data Group} collaboration, M.~Tanabashi et~al., ``{Review of
  Particle
  Physics},''\href{http://dx.doi.org/10.1103/PhysRevD.98.030001}{\emph{Phys.
  Rev. D} {\bf 98} (2018) 030001}.

\bibitem{Sakharov:1967dj}
A.~Sakharov, ``{Violation of CP Invariance, C asymmetry, and baryon asymmetry
  of the
  universe},''\href{http://dx.doi.org/10.1070/PU1991v034n05ABEH002497}{\emph{Sov.
  Phys. Usp.} {\bf 34} (1991) 392--393}.

\bibitem{Kuzmin:1970nx}
V.~Kuzmin, ``{Cp violation and baryon asymmetry of the universe},''{\emph{Pisma
  Zh. Eksp. Teor. Fiz.} {\bf 12} (1970) 335--337}.

\bibitem{Ignatiev:1978uf}
A.~Ignatiev, N.~Krasnikov, V.~Kuzmin and A.~Tavkhelidze, ``{Universal CP
  Noninvariant Superweak Interaction and Baryon Asymmetry of the
  Universe},''\href{http://dx.doi.org/10.1016/0370-2693(78)90900-0}{\emph{Phys.
  Lett. B} {\bf 76} (1978) 436--438}.

\bibitem{Rummukainen:1998as}
K.~Rummukainen, M.~Tsypin, K.~Kajantie, M.~Laine and M.~E. Shaposhnikov, ``{The
  Universality class of the electroweak
  theory},''\href{http://dx.doi.org/10.1016/S0550-3213(98)00494-5}{\emph{Nucl.
  Phys. B} {\bf 532} (1998) 283--314},
  [\href{https://arxiv.org/abs/hep-lat/9805013}{{\tt hep-lat/9805013}}].

\bibitem{Bodeker:2020ghk}
D.~Bodeker and W.~Buchmuller, ``{Baryogenesis from the weak scale to the GUT
  scale},'' \href{https://arxiv.org/abs/2009.07294}{{\tt 2009.07294}}.

\bibitem{Morrissey:2012db}
D.~E. Morrissey and M.~J. Ramsey-Musolf, ``{Electroweak
  baryogenesis},''\href{http://dx.doi.org/10.1088/1367-2630/14/12/125003}{\emph{New
  J. Phys.} {\bf 14} (2012) 125003},
  [\href{https://arxiv.org/abs/1206.2942}{{\tt 1206.2942}}].

\bibitem{Cline:2006ts}
J.~M. Cline, ``{Baryogenesis},'' in \emph{{Les Houches Summer School - Session
  86: Particle Physics and Cosmology: The Fabric of Spacetime}}, 9, 2006.
\newblock \href{https://arxiv.org/abs/hep-ph/0609145}{{\tt hep-ph/0609145}}.

\bibitem{Trodden:1998ym}
M.~Trodden, ``{Electroweak
  baryogenesis},''\href{http://dx.doi.org/10.1103/RevModPhys.71.1463}{\emph{Rev.
  Mod. Phys.} {\bf 71} (1999) 1463--1500},
  [\href{https://arxiv.org/abs/hep-ph/9803479}{{\tt hep-ph/9803479}}].

\bibitem{Arkani-Hamed:2015vfh}
N.~Arkani-Hamed, T.~Han, M.~Mangano and L.-T. Wang, ``{Physics opportunities of
  a 100 TeV proton--proton
  collider},''\href{http://dx.doi.org/10.1016/j.physrep.2016.07.004}{\emph{Phys.
  Rept.} {\bf 652} (2016) 1--49}, [\href{https://arxiv.org/abs/1511.06495}{{\tt
  1511.06495}}].

\bibitem{Mazumdar:2018dfl}
A.~Mazumdar and G.~White, ``{Review of cosmic phase transitions: their
  significance and experimental
  signatures},''\href{http://dx.doi.org/10.1088/1361-6633/ab1f55}{\emph{Rept.
  Prog. Phys.} {\bf 82} (2019) 076901},
  [\href{https://arxiv.org/abs/1811.01948}{{\tt 1811.01948}}].

\bibitem{Joyce:1994fu}
M.~Joyce, T.~Prokopec and N.~Turok, ``{Electroweak baryogenesis from a
  classical
  force},''\href{http://dx.doi.org/10.1103/PhysRevLett.75.1695}{\emph{Phys.
  Rev. Lett.} {\bf 75} (1995) 1695--1698},
  [\href{https://arxiv.org/abs/hep-ph/9408339}{{\tt hep-ph/9408339}}].

\bibitem{Joyce:1994zt}
M.~Joyce, T.~Prokopec and N.~Turok, ``{Nonlocal electroweak baryogenesis. Part
  2: The Classical
  regime},''\href{http://dx.doi.org/10.1103/PhysRevD.53.2958}{\emph{Phys. Rev.
  D} {\bf 53} (1996) 2958--2980},
  [\href{https://arxiv.org/abs/hep-ph/9410282}{{\tt hep-ph/9410282}}].

\bibitem{Fromme:2006wx}
L.~Fromme and S.~J. Huber, ``{Top transport in electroweak
  baryogenesis},''\href{http://dx.doi.org/10.1088/1126-6708/2007/03/049}{\emph{JHEP}
  {\bf 03} (2007) 049}, [\href{https://arxiv.org/abs/hep-ph/0604159}{{\tt
  hep-ph/0604159}}].

\bibitem{Jiang:2015cwa}
M.~Jiang, L.~Bian, W.~Huang and J.~Shu, ``{Impact of a complex singlet:
  Electroweak baryogenesis and dark
  matter},''\href{http://dx.doi.org/10.1103/PhysRevD.93.065032}{\emph{Phys.
  Rev. D} {\bf 93} (2016) 065032},
  [\href{https://arxiv.org/abs/1502.07574}{{\tt 1502.07574}}].

\bibitem{Bruggisser:2018mus}
S.~Bruggisser, B.~Von~Harling, O.~Matsedonskyi and G.~Servant, ``{Baryon
  Asymmetry from a Composite Higgs
  Boson},''\href{http://dx.doi.org/10.1103/PhysRevLett.121.131801}{\emph{Phys.
  Rev. Lett.} {\bf 121} (2018) 131801},
  [\href{https://arxiv.org/abs/1803.08546}{{\tt 1803.08546}}].

\bibitem{Chao:2019smr}
W.~Chao and Y.~Liu, ``{CP violation in the top-assisted electroweak
  baryogenesis},'' \href{https://arxiv.org/abs/1910.09303}{{\tt 1910.09303}}.

\bibitem{Ellis:2019flb}
S.~A. Ellis, S.~Ipek and G.~White, ``{Electroweak Baryogenesis from
  Temperature-Varying
  Couplings},''\href{http://dx.doi.org/10.1007/JHEP08(2019)002}{\emph{JHEP}
  {\bf 08} (2019) 002}, [\href{https://arxiv.org/abs/1905.11994}{{\tt
  1905.11994}}].

\bibitem{DeCurtis:2019rxl}
S.~De~Curtis, L.~Delle~Rose and G.~Panico, ``{Composite Dynamics in the Early
  Universe},''\href{http://dx.doi.org/10.1007/JHEP12(2019)149}{\emph{JHEP} {\bf
  12} (2019) 149}, [\href{https://arxiv.org/abs/1909.07894}{{\tt 1909.07894}}].

\bibitem{Cline:2020jre}
J.~M. Cline and K.~Kainulainen, ``{Electroweak baryogenesis at high bubble wall
  velocities},''\href{http://dx.doi.org/10.1103/PhysRevD.101.063525}{\emph{Phys.
  Rev. D} {\bf 101} (2020) 063525},
  [\href{https://arxiv.org/abs/2001.00568}{{\tt 2001.00568}}].

\bibitem{Xie:2020bkl}
K.-P. Xie, Y.~Wu and L.~Bian, ``{Electroweak baryogenesis and gravitational
  waves in a composite Higgs model with high dimensional fermion
  representations},'' \href{https://arxiv.org/abs/2005.13552}{{\tt
  2005.13552}}.

\bibitem{Modak:2018csw}
T.~Modak and E.~Senaha, ``{Electroweak baryogenesis via bottom
  transport},''\href{http://dx.doi.org/10.1103/PhysRevD.99.115022}{\emph{Phys.
  Rev. D} {\bf 99} (2019) 115022},
  [\href{https://arxiv.org/abs/1811.08088}{{\tt 1811.08088}}].

\bibitem{Modak:2020uyq}
T.~Modak and E.~Senaha, ``{Probing Electroweak Baryogenesis induced by extra
  bottom Yukawa coupling in $bg\to bA \to b t \bar t$ signature},''
  \href{https://arxiv.org/abs/2005.09928}{{\tt 2005.09928}}.

\bibitem{Bruggisser:2018mrt}
S.~Bruggisser, B.~Von~Harling, O.~Matsedonskyi and G.~Servant, ``{Electroweak
  Phase Transition and Baryogenesis in Composite Higgs
  Models},''\href{http://dx.doi.org/10.1007/JHEP12(2018)099}{\emph{JHEP} {\bf
  12} (2018) 099}, [\href{https://arxiv.org/abs/1804.07314}{{\tt 1804.07314}}].

\bibitem{deVries:2018tgs}
J.~De~Vries, M.~Postma and J.~van~de Vis, ``{The role of leptons in electroweak
  baryogenesis},''\href{http://dx.doi.org/10.1007/JHEP04(2019)024}{\emph{JHEP}
  {\bf 04} (2019) 024}, [\href{https://arxiv.org/abs/1811.11104}{{\tt
  1811.11104}}].

\bibitem{Joyce:1994bi}
M.~Joyce, T.~Prokopec and N.~Turok, ``{Efficient electroweak baryogenesis from
  lepton
  transport},''\href{http://dx.doi.org/10.1016/0370-2693(94)91377-3}{\emph{Phys.
  Lett. B} {\bf 338} (1994) 269--275},
  [\href{https://arxiv.org/abs/hep-ph/9401352}{{\tt hep-ph/9401352}}].

\bibitem{Cirigliano:2016nyn}
V.~Cirigliano, W.~Dekens, J.~de~Vries and E.~Mereghetti, ``{Constraining the
  top-Higgs sector of the Standard Model Effective Field
  Theory},''\href{http://dx.doi.org/10.1103/PhysRevD.94.034031}{\emph{Phys.
  Rev. D} {\bf 94} (2016) 034031},
  [\href{https://arxiv.org/abs/1605.04311}{{\tt 1605.04311}}].

\bibitem{deVries:2017ncy}
J.~de~Vries, M.~Postma, J.~van~de Vis and G.~White, ``{Electroweak Baryogenesis
  and the Standard Model Effective Field
  Theory},''\href{http://dx.doi.org/10.1007/JHEP01(2018)089}{\emph{JHEP} {\bf
  01} (2018) 089}, [\href{https://arxiv.org/abs/1710.04061}{{\tt 1710.04061}}].

\bibitem{Espinosa:2011eu}
J.~R. Espinosa, B.~Gripaios, T.~Konstandin and F.~Riva, ``{Electroweak
  Baryogenesis in Non-minimal Composite Higgs
  Models},''\href{http://dx.doi.org/10.1088/1475-7516/2012/01/012}{\emph{JCAP}
  {\bf 01} (2012) 012}, [\href{https://arxiv.org/abs/1110.2876}{{\tt
  1110.2876}}].

\bibitem{Giudice:1993bb}
G.~Giudice and M.~E. Shaposhnikov, ``{Strong sphalerons and electroweak
  baryogenesis},''\href{http://dx.doi.org/10.1016/0370-2693(94)91202-5}{\emph{Phys.
  Lett. B} {\bf 326} (1994) 118--124},
  [\href{https://arxiv.org/abs/hep-ph/9311367}{{\tt hep-ph/9311367}}].

\bibitem{Tulin:2011wi}
S.~Tulin and P.~Winslow, ``{Anomalous $B$ meson mixing and
  baryogenesis},''\href{http://dx.doi.org/10.1103/PhysRevD.84.034013}{\emph{Phys.
  Rev. D} {\bf 84} (2011) 034013}, [\href{https://arxiv.org/abs/1105.2848}{{\tt
  1105.2848}}].

\bibitem{Brod:2013cka}
J.~Brod, U.~Haisch and J.~Zupan, ``{Constraints on CP-violating Higgs couplings
  to the third
  generation},''\href{http://dx.doi.org/10.1007/JHEP11(2013)180}{\emph{JHEP}
  {\bf 11} (2013) 180}, [\href{https://arxiv.org/abs/1310.1385}{{\tt
  1310.1385}}].

\bibitem{Joyce:1994zn}
M.~Joyce, T.~Prokopec and N.~Turok, ``{Nonlocal electroweak baryogenesis. Part
  1: Thin wall
  regime},''\href{http://dx.doi.org/10.1103/PhysRevD.53.2930}{\emph{Phys. Rev.
  D} {\bf 53} (1996) 2930--2957},
  [\href{https://arxiv.org/abs/hep-ph/9410281}{{\tt hep-ph/9410281}}].

\bibitem{Chung:2008aya}
D.~J. Chung, B.~Garbrecht, M.~J. Ramsey-Musolf and S.~Tulin, ``{Yukawa
  Interactions and Supersymmetric Electroweak
  Baryogenesis},''\href{http://dx.doi.org/10.1103/PhysRevLett.102.061301}{\emph{Phys.
  Rev. Lett.} {\bf 102} (2009) 061301},
  [\href{https://arxiv.org/abs/0808.1144}{{\tt 0808.1144}}].

\bibitem{Chung:2009cb}
D.~J. Chung, B.~Garbrecht, M.~J. Ramsey-Musolf and S.~Tulin, ``{Lepton-mediated
  electroweak
  baryogenesis},''\href{http://dx.doi.org/10.1103/PhysRevD.81.063506}{\emph{Phys.
  Rev. D} {\bf 81} (2010) 063506}, [\href{https://arxiv.org/abs/0905.4509}{{\tt
  0905.4509}}].

\bibitem{Chiang:2016vgf}
C.-W. Chiang, K.~Fuyuto and E.~Senaha, ``{Electroweak Baryogenesis with Lepton
  Flavor
  Violation},''\href{http://dx.doi.org/10.1016/j.physletb.2016.09.052}{\emph{Phys.
  Lett. B} {\bf 762} (2016) 315--320},
  [\href{https://arxiv.org/abs/1607.07316}{{\tt 1607.07316}}].

\bibitem{Guo:2016ixx}
H.-K. Guo, Y.-Y. Li, T.~Liu, M.~Ramsey-Musolf and J.~Shu, ``{Lepton-Flavored
  Electroweak
  Baryogenesis},''\href{http://dx.doi.org/10.1103/PhysRevD.96.115034}{\emph{Phys.
  Rev. D} {\bf 96} (2017) 115034},
  [\href{https://arxiv.org/abs/1609.09849}{{\tt 1609.09849}}].

\bibitem{Fuchs:2020uoc}
E.~Fuchs, M.~Losada, Y.~Nir and Y.~Viernik, ``{$CP$ violation from $\tau$, $t$
  and $b$ dimension-6 Yukawa couplings - interplay of baryogenesis, EDM and
  Higgs physics},''\href{http://dx.doi.org/10.1007/JHEP05(2020)056}{\emph{JHEP}
  {\bf 05} (2020) 056}, [\href{https://arxiv.org/abs/2003.00099}{{\tt
  2003.00099}}].

\bibitem{Fuchs:2019ore}
E.~Fuchs, M.~Losada, Y.~Nir and Y.~Viernik, ``{Implications of the Upper Bound
  on $\boldsymbol{h\to\mu^+\mu^-}$ on the Baryon Asymmetry of the
  Universe},''\href{http://dx.doi.org/10.1103/PhysRevLett.124.181801}{\emph{Phys.
  Rev. Lett.} {\bf 124} (2020) 181801},
  [\href{https://arxiv.org/abs/1911.08495}{{\tt 1911.08495}}].

\bibitem{Fernandez-Martinez:2020szk}
E.~Fern\'andez-Mart\'\i{}nez, J.~L\'opez-Pav\'on, T.~Ota and
  S.~Rosauro-Alcaraz, ``{$\nu$ electroweak
  baryogenesis},''\href{http://dx.doi.org/10.1007/JHEP10(2020)063}{\emph{JHEP}
  {\bf 10} (2020) 063}, [\href{https://arxiv.org/abs/2007.11008}{{\tt
  2007.11008}}].

\bibitem{McDonald:1993ey}
J.~McDonald, ``{Electroweak baryogenesis and dark matter via a gauge singlet
  scalar},''\href{http://dx.doi.org/10.1016/0370-2693(94)91229-7}{\emph{Phys.
  Lett. B} {\bf 323} (1994) 339--346}.

\bibitem{Profumo:2007wc}
S.~Profumo, M.~J. Ramsey-Musolf and G.~Shaughnessy, ``{Singlet Higgs
  phenomenology and the electroweak phase
  transition},''\href{http://dx.doi.org/10.1088/1126-6708/2007/08/010}{\emph{JHEP}
  {\bf 08} (2007) 010}, [\href{https://arxiv.org/abs/0705.2425}{{\tt
  0705.2425}}].

\bibitem{Espinosa:2011ax}
J.~R. Espinosa, T.~Konstandin and F.~Riva, ``{Strong Electroweak Phase
  Transitions in the Standard Model with a
  Singlet},''\href{http://dx.doi.org/10.1016/j.nuclphysb.2011.09.010}{\emph{Nucl.
  Phys. B} {\bf 854} (2012) 592--630},
  [\href{https://arxiv.org/abs/1107.5441}{{\tt 1107.5441}}].

\bibitem{Cline:2012hg}
J.~M. Cline and K.~Kainulainen, ``{Electroweak baryogenesis and dark matter
  from a singlet
  Higgs},''\href{http://dx.doi.org/10.1088/1475-7516/2013/01/012}{\emph{JCAP}
  {\bf 01} (2013) 012}, [\href{https://arxiv.org/abs/1210.4196}{{\tt
  1210.4196}}].

\bibitem{Alanne:2014bra}
T.~Alanne, K.~Tuominen and V.~Vaskonen, ``{Strong phase transition, dark matter
  and vacuum stability from simple hidden
  sectors},''\href{http://dx.doi.org/10.1016/j.nuclphysb.2014.11.001}{\emph{Nucl.
  Phys. B} {\bf 889} (2014) 692--711},
  [\href{https://arxiv.org/abs/1407.0688}{{\tt 1407.0688}}].

\bibitem{Vaskonen:2016yiu}
V.~Vaskonen, ``{Electroweak baryogenesis and gravitational waves from a real
  scalar
  singlet},''\href{http://dx.doi.org/10.1103/PhysRevD.95.123515}{\emph{Phys.
  Rev. D} {\bf 95} (2017) 123515},
  [\href{https://arxiv.org/abs/1611.02073}{{\tt 1611.02073}}].

\bibitem{Huang:2018aja}
F.~P. Huang, Z.~Qian and M.~Zhang, ``{Exploring dynamical CP violation induced
  baryogenesis by gravitational waves and
  colliders},''\href{http://dx.doi.org/10.1103/PhysRevD.98.015014}{\emph{Phys.
  Rev. D} {\bf 98} (2018) 015014},
  [\href{https://arxiv.org/abs/1804.06813}{{\tt 1804.06813}}].

\bibitem{Cheng:2018ajh}
W.~Cheng and L.~Bian, ``{From inflation to cosmological electroweak phase
  transition with a complex scalar
  singlet},''\href{http://dx.doi.org/10.1103/PhysRevD.98.023524}{\emph{Phys.
  Rev. D} {\bf 98} (2018) 023524},
  [\href{https://arxiv.org/abs/1801.00662}{{\tt 1801.00662}}].

\bibitem{Alanne:2019bsm}
T.~Alanne, T.~Hugle, M.~Platscher and K.~Schmitz, ``{A fresh look at the
  gravitational-wave signal from cosmological phase
  transitions},''\href{http://dx.doi.org/10.1007/JHEP03(2020)004}{\emph{JHEP}
  {\bf 03} (2020) 004}, [\href{https://arxiv.org/abs/1909.11356}{{\tt
  1909.11356}}].

\bibitem{Gould:2019qek}
O.~Gould, J.~Kozaczuk, L.~Niemi, M.~J. Ramsey-Musolf, T.~V. Tenkanen and D.~J.
  Weir, ``{Nonperturbative analysis of the gravitational waves from a
  first-order electroweak phase
  transition},''\href{http://dx.doi.org/10.1103/PhysRevD.100.115024}{\emph{Phys.
  Rev. D} {\bf 100} (2019) 115024},
  [\href{https://arxiv.org/abs/1903.11604}{{\tt 1903.11604}}].

\bibitem{Carena:2019une}
M.~Carena, Z.~Liu and Y.~Wang, ``{Electroweak phase transition with spontaneous
  Z$_{2}$-breaking},''\href{http://dx.doi.org/10.1007/JHEP08(2020)107}{\emph{JHEP}
  {\bf 08} (2020) 107}, [\href{https://arxiv.org/abs/1911.10206}{{\tt
  1911.10206}}].

\bibitem{Bian:2019kmg}
L.~Bian, Y.~Wu and K.-P. Xie, ``{Electroweak phase transition with composite
  Higgs models: calculability, gravitational waves and collider
  searches},''\href{http://dx.doi.org/10.1007/JHEP12(2019)028}{\emph{JHEP} {\bf
  12} (2019) 028}, [\href{https://arxiv.org/abs/1909.02014}{{\tt 1909.02014}}].

\bibitem{Dolan:1973qd}
L.~Dolan and R.~Jackiw, ``{Symmetry Behavior at Finite
  Temperature},''\href{http://dx.doi.org/10.1103/PhysRevD.9.3320}{\emph{Phys.
  Rev. D} {\bf 9} (1974) 3320--3341}.

\bibitem{Linde:1981zj}
A.~D. Linde, ``{Decay of the False Vacuum at Finite
  Temperature},''\href{http://dx.doi.org/10.1016/0550-3213(83)90072-X}{\emph{Nucl.
  Phys. B} {\bf 216} (1983) 421}.

\bibitem{Quiros:1999jp}
M.~Quiros, ``{Finite temperature field theory and phase transitions},'' in
  \emph{{ICTP Summer School in High-Energy Physics and Cosmology}},
  pp.~187--259, 1, 1999.
\newblock \href{https://arxiv.org/abs/hep-ph/9901312}{{\tt hep-ph/9901312}}.

\bibitem{Moore:1998swa}
G.~D. Moore, ``{Measuring the broken phase sphaleron rate
  nonperturbatively},''\href{http://dx.doi.org/10.1103/PhysRevD.59.014503}{\emph{Phys.
  Rev. D} {\bf 59} (1999) 014503},
  [\href{https://arxiv.org/abs/hep-ph/9805264}{{\tt hep-ph/9805264}}].

\bibitem{Zhou:2019uzq}
R.~Zhou, L.~Bian and H.-K. Guo, ``{Connecting the electroweak sphaleron with
  gravitational
  waves},''\href{http://dx.doi.org/10.1103/PhysRevD.101.091903}{\emph{Phys.
  Rev. D} {\bf 101} (2020) 091903},
  [\href{https://arxiv.org/abs/1910.00234}{{\tt 1910.00234}}].

\bibitem{Wainwright:2011kj}
C.~L. Wainwright, ``{CosmoTransitions: Computing Cosmological Phase Transition
  Temperatures and Bubble Profiles with Multiple
  Fields},''\href{http://dx.doi.org/10.1016/j.cpc.2012.04.004}{\emph{Comput.
  Phys. Commun.} {\bf 183} (2012) 2006--2013},
  [\href{https://arxiv.org/abs/1109.4189}{{\tt 1109.4189}}].

\bibitem{Audley:2017drz}
{\scshape LISA} collaboration, P.~Amaro-Seoane et~al., ``{Laser Interferometer
  Space Antenna},'' \href{https://arxiv.org/abs/1702.00786}{{\tt 1702.00786}}.

\bibitem{Crowder:2005nr}
J.~Crowder and N.~J. Cornish, ``{Beyond LISA: Exploring future gravitational
  wave
  missions},''\href{http://dx.doi.org/10.1103/PhysRevD.72.083005}{\emph{Phys.
  Rev. D} {\bf 72} (2005) 083005},
  [\href{https://arxiv.org/abs/gr-qc/0506015}{{\tt gr-qc/0506015}}].

\bibitem{Luo:2015ght}
{\scshape TianQin} collaboration, J.~Luo et~al., ``{TianQin: a space-borne
  gravitational wave
  detector},''\href{http://dx.doi.org/10.1088/0264-9381/33/3/035010}{\emph{Class.
  Quant. Grav.} {\bf 33} (2016) 035010},
  [\href{https://arxiv.org/abs/1512.02076}{{\tt 1512.02076}}].

\bibitem{Hu:2017yoc}
Y.-M. Hu, J.~Mei and J.~Luo, ``{Science prospects for space-borne
  gravitational-wave
  missions},''\href{http://dx.doi.org/10.1093/nsr/nwx115}{\emph{Natl. Sci.
  Rev.} {\bf 4} (2017) 683--684}.

\bibitem{Hu:2017mde}
W.-R. Hu and Y.-L. Wu, ``{The Taiji Program in Space for gravitational wave
  physics and the nature of
  gravity},''\href{http://dx.doi.org/10.1093/nsr/nwx116}{\emph{Natl. Sci. Rev.}
  {\bf 4} (2017) 685--686}.

\bibitem{Guo:2018npi}
W.-H. Ruan, Z.-K. Guo, R.-G. Cai and Y.-Z. Zhang, ``{Taiji program:
  Gravitational-wave
  sources},''\href{http://dx.doi.org/10.1142/S0217751X2050075X}{\emph{Int. J.
  Mod. Phys. A} {\bf 35} (2020) 2050075},
  [\href{https://arxiv.org/abs/1807.09495}{{\tt 1807.09495}}].

\bibitem{Kawamura:2011zz}
S.~Kawamura et~al., ``{The Japanese space gravitational wave antenna:
  DECIGO},''\href{http://dx.doi.org/10.1088/0264-9381/28/9/094011}{\emph{Class.
  Quant. Grav.} {\bf 28} (2011) 094011}.

\bibitem{Kawamura:2006up}
S.~Kawamura et~al., ``{The Japanese space gravitational wave antenna
  DECIGO},''\href{http://dx.doi.org/10.1088/0264-9381/23/8/S17}{\emph{Class.
  Quant. Grav.} {\bf 23} (2006) S125--S132}.

\bibitem{Ellis:2018mja}
J.~Ellis, M.~Lewicki and J.~M. No, ``{On the Maximal Strength of a First-Order
  Electroweak Phase Transition and its Gravitational Wave
  Signal},''\href{http://dx.doi.org/10.1088/1475-7516/2019/04/003}{\emph{JCAP}
  {\bf 04} (2019) 003}, [\href{https://arxiv.org/abs/1809.08242}{{\tt
  1809.08242}}].

\bibitem{Grojean:2006bp}
C.~Grojean and G.~Servant, ``{Gravitational Waves from Phase Transitions at the
  Electroweak Scale and
  Beyond},''\href{http://dx.doi.org/10.1103/PhysRevD.75.043507}{\emph{Phys.
  Rev. D} {\bf 75} (2007) 043507},
  [\href{https://arxiv.org/abs/hep-ph/0607107}{{\tt hep-ph/0607107}}].

\bibitem{Caprini:2015zlo}
C.~Caprini et~al., ``{Science with the space-based interferometer eLISA. II:
  Gravitational waves from cosmological phase
  transitions},''\href{http://dx.doi.org/10.1088/1475-7516/2016/04/001}{\emph{JCAP}
  {\bf 04} (2016) 001}, [\href{https://arxiv.org/abs/1512.06239}{{\tt
  1512.06239}}].

\bibitem{Caprini:2019egz}
C.~Caprini et~al., ``{Detecting gravitational waves from cosmological phase
  transitions with LISA: an
  update},''\href{http://dx.doi.org/10.1088/1475-7516/2020/03/024}{\emph{JCAP}
  {\bf 03} (2020) 024}, [\href{https://arxiv.org/abs/1910.13125}{{\tt
  1910.13125}}].

\bibitem{Espinosa:2010hh}
J.~R. Espinosa, T.~Konstandin, J.~M. No and G.~Servant, ``{Energy Budget of
  Cosmological First-order Phase
  Transitions},''\href{http://dx.doi.org/10.1088/1475-7516/2010/06/028}{\emph{JCAP}
  {\bf 06} (2010) 028}, [\href{https://arxiv.org/abs/1004.4187}{{\tt
  1004.4187}}].

\bibitem{Wang:2020nzm}
X.~Wang, F.~P. Huang and X.~Zhang, ``{The energy budget and the gravitational
  wave spectra beyond the bag model},''
  \href{https://arxiv.org/abs/2010.13770}{{\tt 2010.13770}}.

\bibitem{Megevand:2016lpr}
A.~Megevand and S.~Ramirez, ``{Bubble nucleation and growth in very strong
  cosmological phase
  transitions},''\href{http://dx.doi.org/10.1016/j.nuclphysb.2017.03.009}{\emph{Nucl.
  Phys. B} {\bf 919} (2017) 74--109},
  [\href{https://arxiv.org/abs/1611.05853}{{\tt 1611.05853}}].

\bibitem{Kobakhidze:2017mru}
A.~Kobakhidze, C.~Lagger, A.~Manning and J.~Yue, ``{Gravitational waves from a
  supercooled electroweak phase transition and their detection with pulsar
  timing
  arrays},''\href{http://dx.doi.org/10.1140/epjc/s10052-017-5132-y}{\emph{Eur.
  Phys. J. C} {\bf 77} (2017) 570},
  [\href{https://arxiv.org/abs/1703.06552}{{\tt 1703.06552}}].

\bibitem{Ellis:2020awk}
J.~Ellis, M.~Lewicki and J.~M. No, ``{Gravitational waves from first-order
  cosmological phase transitions: lifetime of the sound wave
  source},''\href{http://dx.doi.org/10.1088/1475-7516/2020/07/050}{\emph{JCAP}
  {\bf 07} (2020) 050}, [\href{https://arxiv.org/abs/2003.07360}{{\tt
  2003.07360}}].

\bibitem{Wang:2020jrd}
X.~Wang, F.~P. Huang and X.~Zhang, ``{Phase transition dynamics and
  gravitational wave spectra of strong first-order phase transition in
  supercooled
  universe},''\href{http://dx.doi.org/10.1088/1475-7516/2020/05/045}{\emph{JCAP}
  {\bf 05} (2020) 045}, [\href{https://arxiv.org/abs/2003.08892}{{\tt
  2003.08892}}].

\bibitem{Guo:2020grp}
H.-K. Guo, K.~Sinha, D.~Vagie and G.~White, ``{Phase Transitions in an
  Expanding Universe: Stochastic Gravitational Waves in Standard and
  Non-Standard Histories},'' \href{https://arxiv.org/abs/2007.08537}{{\tt
  2007.08537}}.

\bibitem{No:2011fi}
J.~M. No, ``{Large Gravitational Wave Background Signals in Electroweak
  Baryogenesis
  Scenarios},''\href{http://dx.doi.org/10.1103/PhysRevD.84.124025}{\emph{Phys.
  Rev. D} {\bf 84} (2011) 124025}, [\href{https://arxiv.org/abs/1103.2159}{{\tt
  1103.2159}}].

\bibitem{Lee:2004we}
C.~Lee, V.~Cirigliano and M.~J. Ramsey-Musolf, ``{Resonant relaxation in
  electroweak
  baryogenesis},''\href{http://dx.doi.org/10.1103/PhysRevD.71.075010}{\emph{Phys.
  Rev. D} {\bf 71} (2005) 075010},
  [\href{https://arxiv.org/abs/hep-ph/0412354}{{\tt hep-ph/0412354}}].

\bibitem{DOnofrio:2014rug}
M.~D'Onofrio, K.~Rummukainen and A.~Tranberg, ``{Sphaleron Rate in the Minimal
  Standard
  Model},''\href{http://dx.doi.org/10.1103/PhysRevLett.113.141602}{\emph{Phys.
  Rev. Lett.} {\bf 113} (2014) 141602},
  [\href{https://arxiv.org/abs/1404.3565}{{\tt 1404.3565}}].

\bibitem{Konstandin:2014zta}
T.~Konstandin, G.~Nardini and I.~Rues, ``{From Boltzmann equations to steady
  wall
  velocities},''\href{http://dx.doi.org/10.1088/1475-7516/2014/09/028}{\emph{JCAP}
  {\bf 09} (2014) 028}, [\href{https://arxiv.org/abs/1407.3132}{{\tt
  1407.3132}}].

\bibitem{Profumo:2014opa}
S.~Profumo, M.~J. Ramsey-Musolf, C.~L. Wainwright and P.~Winslow,
  ``{Singlet-catalyzed electroweak phase transitions and precision Higgs boson
  studies},''\href{http://dx.doi.org/10.1103/PhysRevD.91.035018}{\emph{Phys.
  Rev. D} {\bf 91} (2015) 035018}, [\href{https://arxiv.org/abs/1407.5342}{{\tt
  1407.5342}}].

\bibitem{Cao:2017oez}
Q.-H. Cao, F.~P. Huang, K.-P. Xie and X.~Zhang, ``{Testing the electroweak
  phase transition in scalar extension models at lepton
  colliders},''\href{http://dx.doi.org/10.1088/1674-1137/42/2/023103}{\emph{Chin.
  Phys. C} {\bf 42} (2018) 023103},
  [\href{https://arxiv.org/abs/1708.04737}{{\tt 1708.04737}}].

\bibitem{Alves:2018oct}
A.~Alves, T.~Ghosh, H.-K. Guo and K.~Sinha, ``{Resonant Di-Higgs Production at
  Gravitational Wave Benchmarks: A Collider Study using Machine
  Learning},''\href{http://dx.doi.org/10.1007/JHEP12(2018)070}{\emph{JHEP} {\bf
  12} (2018) 070}, [\href{https://arxiv.org/abs/1808.08974}{{\tt 1808.08974}}].

\bibitem{Zhou:2020idp}
L.~Bian, H.-K. Guo, Y.~Wu and R.~Zhou, ``{Gravitational wave and collider
  searches for electroweak symmetry breaking
  patterns},''\href{http://dx.doi.org/10.1103/PhysRevD.101.035011}{\emph{Phys.
  Rev. D} {\bf 101} (2020) 035011},
  [\href{https://arxiv.org/abs/1906.11664}{{\tt 1906.11664}}].

\bibitem{Chen:2019ebq}
N.~Chen, T.~Li, Y.~Wu and L.~Bian, ``{Complementarity of the future $e^+ e^-$
  colliders and gravitational waves in the probe of complex singlet extension
  to the standard
  model},''\href{http://dx.doi.org/10.1103/PhysRevD.101.075047}{\emph{Phys.
  Rev. D} {\bf 101} (2020) 075047},
  [\href{https://arxiv.org/abs/1911.05579}{{\tt 1911.05579}}].

\bibitem{Huang:2016cjm}
P.~Huang, A.~J. Long and L.-T. Wang, ``{Probing the Electroweak Phase
  Transition with Higgs Factories and Gravitational
  Waves},''\href{http://dx.doi.org/10.1103/PhysRevD.94.075008}{\emph{Phys. Rev.
  D} {\bf 94} (2016) 075008}, [\href{https://arxiv.org/abs/1608.06619}{{\tt
  1608.06619}}].

\bibitem{Kozaczuk:2019pet}
J.~Kozaczuk, M.~J. Ramsey-Musolf and J.~Shelton, ``{Exotic Higgs boson decays
  and the electroweak phase
  transition},''\href{http://dx.doi.org/10.1103/PhysRevD.101.115035}{\emph{Phys.
  Rev. D} {\bf 101} (2020) 115035},
  [\href{https://arxiv.org/abs/1911.10210}{{\tt 1911.10210}}].

\bibitem{Papaefstathiou:2020iag}
A.~Papaefstathiou and G.~White, ``{The Electro-Weak Phase Transition at
  Colliders: Confronting Theoretical Uncertainties and Complementary
  Channels},'' \href{https://arxiv.org/abs/2010.00597}{{\tt 2010.00597}}.

\bibitem{Alves:2020bpi}
A.~Alves, D.~Gon\c{c}alves, T.~Ghosh, H.-K. Guo and K.~Sinha, ``{Di-Higgs Blind
  Spots in Gravitational Wave Signals},''
  \href{https://arxiv.org/abs/2007.15654}{{\tt 2007.15654}}.

\bibitem{Abada:2019ono}
{\scshape FCC} collaboration, A.~Abada et~al., ``{HE-LHC: The High-Energy Large
  Hadron Collider}: {Future Circular Collider Conceptual Design Report Volume
  4},''\href{http://dx.doi.org/10.1140/epjst/e2019-900088-6}{\emph{Eur. Phys.
  J. ST} {\bf 228} (2019) 1109--1382}.

\bibitem{Ashoorioon:2009nf}
A.~Ashoorioon and T.~Konstandin, ``{Strong electroweak phase transitions
  without collider
  traces},''\href{http://dx.doi.org/10.1088/1126-6708/2009/07/086}{\emph{JHEP}
  {\bf 07} (2009) 086}, [\href{https://arxiv.org/abs/0904.0353}{{\tt
  0904.0353}}].

\bibitem{Curtin:2014jma}
D.~Curtin, P.~Meade and C.-T. Yu, ``{Testing Electroweak Baryogenesis with
  Future
  Colliders},''\href{http://dx.doi.org/10.1007/JHEP11(2014)127}{\emph{JHEP}
  {\bf 11} (2014) 127}, [\href{https://arxiv.org/abs/1409.0005}{{\tt
  1409.0005}}].

\bibitem{Bagliesi:2007qx}
G.~Bagliesi, ``{Tau tagging at Atlas and CMS},'' in \emph{{17th Symposium on
  Hadron Collider Physics 2006 (HCP 2006)}}, 7, 2007.
\newblock \href{https://arxiv.org/abs/0707.0928}{{\tt 0707.0928}}.

\bibitem{Aaboud:2018zeb}
{\scshape ATLAS} collaboration, M.~Aaboud et~al., ``{Search for supersymmetry
  in events with four or more leptons in $\sqrt{s}=13$ TeV $pp$ collisions with
  ATLAS},''\href{http://dx.doi.org/10.1103/PhysRevD.98.032009}{\emph{Phys. Rev.
  D} {\bf 98} (2018) 032009}, [\href{https://arxiv.org/abs/1804.03602}{{\tt
  1804.03602}}].

\bibitem{Alloul:2013bka}
A.~Alloul, N.~D. Christensen, C.~Degrande, C.~Duhr and B.~Fuks, ``{FeynRules
  2.0 - A complete toolbox for tree-level
  phenomenology},''\href{http://dx.doi.org/10.1016/j.cpc.2014.04.012}{\emph{Comput.
  Phys. Commun.} {\bf 185} (2014) 2250--2300},
  [\href{https://arxiv.org/abs/1310.1921}{{\tt 1310.1921}}].

\bibitem{Degrande:2011ua}
C.~Degrande, C.~Duhr, B.~Fuks, D.~Grellscheid, O.~Mattelaer and T.~Reiter,
  ``{UFO - The Universal FeynRules
  Output},''\href{http://dx.doi.org/10.1016/j.cpc.2012.01.022}{\emph{Comput.
  Phys. Commun.} {\bf 183} (2012) 1201--1214},
  [\href{https://arxiv.org/abs/1108.2040}{{\tt 1108.2040}}].

\bibitem{Alwall:2014hca}
J.~Alwall, R.~Frederix, S.~Frixione, V.~Hirschi, F.~Maltoni, O.~Mattelaer
  et~al., ``{The automated computation of tree-level and next-to-leading order
  differential cross sections, and their matching to parton shower
  simulations},''\href{http://dx.doi.org/10.1007/JHEP07(2014)079}{\emph{JHEP}
  {\bf 07} (2014) 079}, [\href{https://arxiv.org/abs/1405.0301}{{\tt
  1405.0301}}].

\bibitem{Sjostrand:2007gs}
T.~Sjostrand, S.~Mrenna and P.~Z. Skands, ``{A Brief Introduction to PYTHIA
  8.1},''\href{http://dx.doi.org/10.1016/j.cpc.2008.01.036}{\emph{Comput. Phys.
  Commun.} {\bf 178} (2008) 852--867},
  [\href{https://arxiv.org/abs/0710.3820}{{\tt 0710.3820}}].

\bibitem{deFavereau:2013fsa}
{\scshape DELPHES 3} collaboration, J.~de~Favereau, C.~Delaere, P.~Demin,
  A.~Giammanco, V.~Lema\^\i{}tre, A.~Mertens et~al., ``{DELPHES 3, A modular
  framework for fast simulation of a generic collider
  experiment},''\href{http://dx.doi.org/10.1007/JHEP02(2014)057}{\emph{JHEP}
  {\bf 02} (2014) 057}, [\href{https://arxiv.org/abs/1307.6346}{{\tt
  1307.6346}}].

\bibitem{Aaboud:2018xpj}
{\scshape ATLAS} collaboration, M.~Aaboud et~al., ``{Search for new phenomena
  in events with same-charge leptons and $b$-jets in $pp$ collisions at
  $\sqrt{s}= 13$ TeV with the ATLAS
  detector},''\href{http://dx.doi.org/10.1007/JHEP12(2018)039}{\emph{JHEP} {\bf
  12} (2018) 039}, [\href{https://arxiv.org/abs/1807.11883}{{\tt 1807.11883}}].

\bibitem{Carrington:1991hz}
M.~Carrington, ``{The Effective potential at finite temperature in the Standard
  Model},''\href{http://dx.doi.org/10.1103/PhysRevD.45.2933}{\emph{Phys. Rev.
  D} {\bf 45} (1992) 2933--2944}.

\bibitem{Fuchs:2020pun}
E.~Fuchs, M.~Losada, Y.~Nir and Y.~Viernik, ``{Analytic Techniques for Solving
  the Transport Equations in Electroweak Baryogenesis},''
  \href{https://arxiv.org/abs/2007.06940}{{\tt 2007.06940}}.

\bibitem{Enqvist:1997ff}
K.~Enqvist, A.~Riotto and I.~Vilja, ``{Baryogenesis and the thermalization rate
  of stop},''\href{http://dx.doi.org/10.1016/S0370-2693(98)00963-0}{\emph{Phys.
  Lett. B} {\bf 438} (1998) 273--280},
  [\href{https://arxiv.org/abs/hep-ph/9710373}{{\tt hep-ph/9710373}}].

\bibitem{Elmfors:1998hh}
P.~Elmfors, K.~Enqvist, A.~Riotto and I.~Vilja, ``{Damping rates in the MSSM
  and electroweak
  baryogenesis},''\href{http://dx.doi.org/10.1016/S0370-2693(99)00169-0}{\emph{Phys.
  Lett. B} {\bf 452} (1999) 279--286},
  [\href{https://arxiv.org/abs/hep-ph/9809529}{{\tt hep-ph/9809529}}].

\bibitem{Cirigliano:2006wh}
V.~Cirigliano, M.~J. Ramsey-Musolf, S.~Tulin and C.~Lee, ``{Yukawa and
  tri-scalar processes in electroweak
  baryogenesis},''\href{http://dx.doi.org/10.1103/PhysRevD.73.115009}{\emph{Phys.
  Rev. D} {\bf 73} (2006) 115009},
  [\href{https://arxiv.org/abs/hep-ph/0603058}{{\tt hep-ph/0603058}}].

\end{thebibliography}\endgroup

\end{document}